\documentclass{emulateapj}

\shorttitle{Broad Line Radio Galaxies Observed with {\it Fermi}-LAT}  
\shortauthors{Kataoka et al.}
\usepackage{times}

\def\F{{\it Fermi}-LAT }

\begin{document}

\title{Broad Line Radio Galaxies Observed with {\it Fermi}-LAT:\\
The Origin of the GeV $\gamma$-ray Emission}

\author{
J.~Kataoka\altaffilmark{1,2}, 
{\L}.~Stawarz\altaffilmark{3,4}, 
Y.~Takahashi\altaffilmark{1}, 
C.~C.~Cheung\altaffilmark{5}, 
M.~Hayashida\altaffilmark{6}, 
P.~Grandi\altaffilmark{7}, 
T.~H.~Burnett\altaffilmark{8}, 
A.~Celotti\altaffilmark{9}, 
S.~J.~Fegan\altaffilmark{10}, 
P.~Fortin\altaffilmark{10}, 
K.~Maeda\altaffilmark{1}, 
T.~Nakamori\altaffilmark{1}, 
G.~B.~Taylor\altaffilmark{11}, 
G.~Tosti\altaffilmark{12,13},
S.~W.~Digel\altaffilmark{6},
W.~McConville\altaffilmark{14,15},
J.~Finke\altaffilmark{16},
F.~D'Ammando\altaffilmark{17,18}
}

\altaffiltext{1}{Research Institute for Science and Engineering, Waseda University, 3-4-1, Okubo, Shinjuku, Tokyo 169-8555, Japan}
\altaffiltext{2}{email: kataoka.jun@waseda.jp}
\altaffiltext{3}{Institute of Space and Astronautical Science, JAXA, 3-1-1 Yoshinodai, Chuo-ku, Sagamihara, Kanagawa 252-5210, Japan}
\altaffiltext{4}{Astronomical Observatory, Jagiellonian University, 30-244 Krak\'ow, Poland}
\altaffiltext{5}{National Research Council Research Associate, National Academy of Sciences, Washington, DC 20001, resident at Naval Research Laboratory, Washington, DC 20375, USA}
\altaffiltext{6}{W. W. Hansen Experimental Physics Laboratory, Kavli Institute for Particle Astrophysics and Cosmology, Department of Physics and SLAC National Accelerator Laboratory, Stanford University, Stanford, CA 94305, USA}
\altaffiltext{7}{INAF-IASF Bologna, 40129 Bologna, Italy}
\altaffiltext{8}{Department of Physics, University of Washington, Seattle, WA 98195-1560, USA}
\altaffiltext{9}{Scuola Internazionale Superiore di Studi Avanzati (SISSA), 34014 Trieste, Italy}
\altaffiltext{10}{Laboratoire Leprince-Ringuet, \'Ecole polytechnique, CNRS/IN2P3, Palaiseau, France}
\altaffiltext{11}{University of New Mexico, MSC07 4220, Albuquerque, NM 87131, USA}
\altaffiltext{12}{Istituto Nazionale di Fisica Nucleare, Sezione di Perugia, I-06123 Perugia, Italy}
\altaffiltext{13}{Dipartimento di Fisica, Universit\`a degli Studi di Perugia, I-06123 Perugia, Italy}
\altaffiltext{14}{NASA Goddard Space Flight Center, Greenbelt, MD 20771, USA}
\altaffiltext{15}{Department of Physics and Department of Astronomy,
University of Maryland, College Park, MD 20742, USA}
\altaffiltext{16}{Space Science Division, Naval Research Laboratory,
Washington, DC 20375-5352}
\altaffiltext{17}{IASF Palermo, 90146 Palermo, Italy}
\altaffiltext{18}{INAF-Istituto di Astrofisica Spaziale e Fisica Cosmica, I-00133 Roma, Italy}

\begin{abstract}
We report on a detailed investigation of the $\gamma$-ray emission from 
18 broad line radio galaxies (BLRGs) based on two years of 
{\it Fermi} Large Area Telescope (LAT) 
data. We confirm the 
previously reported detections of 3C~120 and 3C~111 in the GeV 
photon energy range; a detailed look at the temporal characteristics of the 
observed $\gamma$-ray emission reveals in addition possible flux variability 
in both sources. No statistically significant $\gamma$-ray detection of the other BLRGs 
was however found in the considered dataset. Though the 
sample size studied is small, what appears 
to differentiate 3C~111 and 3C~120 from the BLRGs not yet detected 
in $\gamma$-rays is the particularly strong nuclear radio flux. 
This finding, together with the indications of the $\gamma$-ray flux variability
and a number of other arguments presented, 
indicate that the GeV emission of BLRGs is 
most likely dominated by the beamed radiation of 
relativistic jets observed at intermediate viewing angles.
In this paper we also analyzed a comparison sample of high accretion-rate 
Seyfert 1 galaxies, which can be considered
radio-quiet counterparts of BLRGs, and found none were detected in $\gamma$-rays.
A simple phenomenological hybrid model applied for the broad-band 
emission of the discussed radio-loud and radio-quiet type 1 active galaxies 
suggests that the relative contribution of 
the nuclear jets to the accreting matter is $\geq 1\%$ on average 
for BLRGs, whilst $\leq 0.1\%$ for Seyfert 1 galaxies.

\end{abstract}


\section{Introduction}

A long debated problem in our understanding of accreting supermassive
black holes (SMBHs) in the Universe is the unification of different
types of active galactic nuclei (AGN) in a framework 
ascribing their observed diversity to a 
relatively few differing parameters and factors. For
example, it has been widely argued that the difference between type 1 and
type 2 AGN --- i.e., those possessing and lacking broad permitted
emission lines in their nuclear spectra, respectively --- may be
explained by simple geometrical effects involving anisotropic
obscuration of the active center viewed at different inclination with
respect to the accretion disk axis \citep{ant93}. The accretion rate was
claimed to play a role in this context as well, since `standard'
geometrically-thin optically-thick accretion disks formed at high
accretion rates \citep[$> 1\%$ Eddington;][]{sha73} on the one hand,
and radiatively-inefficient geometrically-thick accretion flows expected
to be present at low accretion rates \citep[$< 1\%$
Eddington;][]{nar94,nar95,abr95} on the other hand, can account for
different emission spectra and power outputs of high- and low-power AGN
(see an early discussion on this issue by \citealt{fab95} and
\citealt{las96}, and the more recent one in \citealt{ghi09}). The other
relevant parameter may be also the mass of the SMBH itself, which seems 
to determine some physical differences between Narrow-Line Seyfert 1 
galaxies and `regular' Seyfert 1s \citep[e.g.,][]{pog00,kom08}.

None of the aforementioned factors can, however, account for the
dichotomy between radio-loud AGN (which have jets) and radio-quiet 
AGN (which don't). 
One may therefore seek the fundamental difference between AGN
producing luminous radio-emitting outflows and those lacking such
outflows in yet another parameter of the accretion disk/black hole
system. One possibility is that, for example, radio-loud AGN harbor
rapidly spinning SMBHs with rotational energy extracted
electromagnetically via the \citet{bla77} mechanism, and converted to
the kinetic luminosity of relativistic jets. At the same time, the 
negligible angular momentum of SMBHs hosted by radio-quiet AGN precludes
the formation of such powerful well-collimated outflows. This
so-called `spin paradigm' \citep{bla90} has recently been claimed to be 
supported, after some modifications, by several observational findings
and theoretical investigations \citep[e.g.,][see also in this context
Garofalo 2009]{koi02,sik07,tch10}.
Regardless of this debate, the presence of a relativistic jet
constitutes a fundamental distinction between various types of AGN,
simply because an anisotropic and strongly Doppler-boosted jet emission
can dramatically affect the observed properties of a source. In this
context, geometrical effects play again a major role. In particular, the
total radiative output of those radio-loud AGN which are observed at
small viewing angles to the jet axis (`blazars') may be dominated by the
broad-band jet emission, while those AGN which are inclined at larger
viewing angles (e.g., radio galaxies) may display radiative signatures
of both outflowing and accreting matter at comparable levels \citep[see,
e.g.,][]{bar89,urr95}. We note that in addition to such geometrical
effects, the age of a radio structure (i.e., the time elapsed after the
onset of the jet activity in the nucleus) is yet another factor crucial
to understanding unification of radio-loud AGN \citep[see,
e.g.,][]{odea98}.

Interestingly, new deep radio surveys indicate that the radio-loudness
parameter --- which is defined as a ratio of the jet-related radio flux
to the disk-related optical flux, and is considered as a proxy for the
jet production efficiency\footnote{More accurately, the
radio-loudness parameter depends not only on the total power of a jet
relative to the total power of an accreting matter, but also on
their respective radiative efficiencies. Only if these
radiative efficiencies are not very different between various types of
AGN (and various AGN of the same type), the radio-loudness parameter may
be considered as a good proxy for the jet production efficiency.} ---
shows a continuous distribution rather than a sharp division between
radio-loud and radio-quiet AGN. This applies to AGN hosted by early-type
galaxies and accreting at high rates (i.e., quasars), which are
typically studied in this context \citep[e.g.,][]{whi00}, but holds also
when other elliptical-hosted AGN (radio galaxies) spanning a wide range
of the accretion rate are taken into account \citep{sik07}. Moreover,
unresolved non-thermal radio emission and jet-like structures have been
discovered in classes of AGN considered previously as `radio quiet',
i.e., Seyfert galaxies, although the jets in such systems are
non-relativistic and weak when compared to the jets found in `classical'
radio galaxies and quasars \citep[e.g.,][and references
therein]{ulv89,kuk95,the01,mid04}. The distribution of radio-loudness
parameter in Seyferts, which are typically hosted by late-type (disk)
galaxies is, however, similarly a continuous function of the accretion
rate, as demonstrated first by \citet{ho01} and \citet{ho02}. Yet, it
was pointed out by \citet{sik07} that there is a substantial difference
in the distribution of the jet production efficiency between disk-hosted
and elliptical-hosted AGN, with Seyfert galaxies being characterized, at
any accretion rate, by the radio-loudness parameters orders of
magnitudes smaller than the analogous parameters characterizing
elliptical-hosted radio galaxies or quasars. Clearly, more studies
regarding the jet-disk connection in different types of AGN are needed
to understand the jet launching processes and the physics of active
SMBHs in general.

In this context, broad line radio galaxies (BLRGs) seem ideal targets
for an in-depth investigation, since this particular class of very
radio-loud AGN exhibits both the disk-related (`Seyfert-like') and the
jet-related (`blazar-like') radiative signatures in their broad-band
spectra. Unlike blazars, the jets in BLRGs are not pointing directly
toward the observer, and so the relativistic beaming effects and the
related jet dominance are only moderate. Moreover, unlike narrow-line
radio galaxies (NRLGs) --- which are believed to be intrinsically
similar but simply inclined at systematically larger jet viewing angles --- BLRGs
are not generally obscured by large amounts of dust distributed in
torus-like structures around the nucleus, and hence radiative properties
of the accretion disks and of the circumnuclear gas can be easily
accessed in their case. BLRGs show in particular optical/UV continuum
and emission-line characteristics very similar to those of luminous
Seyfert galaxies, which indicates high accretion rates and the presence of
standard Shakura-Sunyaev disks in both classes of objects. Some
fundamental differences in the X-ray spectra between BLRGs and
high-accretion-rate Seyferts have been however noted. Specifically, even
though the observed X-ray/soft $\gamma$-ray emission of BLRGs seem still
dominated by the moderately absorbed emission by the accreting plasma
(i.e., disks and disk coronae), rather than by the jets, the
$1-100$\,keV continua of BLRGs are flatter, and their reflection
components (as well as the fluorescent Fe K$\alpha$ lines) weaker than
in the case of luminous Seyfert galaxies
\citep[e.g.,][]{mar91,woz98,sam99,sam02,era00,zdz01,bal07}. Because of
such differences, several authors in the past speculated about the
non-negligible jet contribution to the X-ray emission of BLRGs, diluting
the accretion-related radiative output in the X-ray domain
\citep{woz98,era00,gra02}. This idea was subsequently examined in
various different approaches using most recent broadband X-ray data
obtained with {\it BeppoSAX} \citep{gra06,gra07}, {\it Suzaku}
and {\it Swift} \citep{kat07,sam09}.

\citet{gra07} attempted to disentangle the jet and the disk
contributions to the X-ray spectra of three of the brightest BLRGs
\citep[the approach first applied to the case of the quasar 3C~273;
see][]{gra04}. For simplicity, they assumed that the accretion disks in
BLRGs and Seyfert 1 galaxies produce similar emission continua and
reprocessing features, and subsequently allowed for a presence of the
Doppler-enhanced jet radiation at an arbitrary level. The fits obtained
for 3C~120, 3C~390.3 and 3C~382 showed that the data are indeed
consistent with a combination of a thermal component (in a first
approximation associated with an accretion disk) and a non-thermal
component associated with the beamed radiation (due to a
jet). \citeauthor{gra07} concluded however that jets make only minimal
contribution to the X-ray continuum emission of BLRGs, in agreement with
the previous findings by \citet{woz98}. More recently, \citet{sam09}
also proposed that BLRGs may be just clustered at one end of the
distribution of the X-ray spectral parameters (e.g., photon indices
and reflection albedos) characterizing Seyfert galaxies with
significant overlap. Interestingly, both \citet{gra07} and \citet{sam09}
suggested that the emission of the underlying jet may instead dominate
the radiative output of BLRGs at high energy $\gamma$ rays, and in
particular in the GeV regime.

With the successful launch of the {\it Fermi} Gamma-ray Space Telescope,
we now have an unprecedented opportunity to study in detail the
$\gamma$-ray emission from different types of extragalactic sources ---
not only blazars, but also radio galaxies
\citep{PerA,M87,CenA,MAGN,kat10}, and other classes of AGN as
well \citep[such as Narrow-Line Seyfert 1 galaxies, for
example;][]{NLS}. During the first 15-month of the {\it Fermi} mission,
11 non-blazar AGN have been detected in the GeV photon energy range
\citep{FAGN,MAGN} by the {\it Fermi} Large Area Telescope
\citep[LAT;][]{atw09,1FGL}. This `misaligned AGN sample' includes seven
Faranoff-Riley type I (low-power) radio galaxies, and four
Faranoff-Riley type II (high-power, hereafter `FR II') radio sources
consisting of two radio galaxies and two steep spectrum radio
quasars. The two luminous radio galaxies detected in $\gamma$ rays are, 
in fact, X-ray bright objects classified
spectroscopically as BLRGs: 3C~120 (FR I radio morphology), detected 
for the first time with \F, and 3C~111 (FR II radio morphology), 
whose $\gamma$-ray detection was
initially reported by EGRET \citep{har08} and confirmed by \F. 
None of the X-ray bright Seyfert 1 galaxies appear to
be detected in $\gamma$ rays thus far. 

In this paper we report on a detailed investigation of the $\gamma$-ray 
emission from 18 hard X-ray brightest BLRGs, as well as 
a comparison sample of high accretion-rate Seyfert 1 galaxies 
selected as their supposed radio-quiet counterparts (in the framework 
of the AGN unification scheme). Our primary goals are
to examine the $\gamma$-ray properties of BLRGs as potential 
`$\gamma$-ray loud' active galactic nuclei, to study the high-energy
jet emission in the selected sources, in particular, investigating the relative 
contributions of the nuclear jet and accretion disk emission  
in the broad-band spectra of BLRGs. The two
years of \F exposure provides us with a rather deep all-sky survey
reaching the flux limit of typically a few $\times
10^{-12}$\,erg\,cm$^{-2}$\,s$^{-1}$ ($95\%$ confidence level) between
the observed photon energies 100\,MeV and 10\,GeV. In $\S$\,2, we
describe the \F observations and data reduction procedure. The results
of the analysis are given in $\S$\,3. The discussion and final
conclusions are presented in $\S$\,4. Throughout this paper, a
$\Lambda$CDM cosmology with $H_0 = 71$\,km\,s$^{-1}$\,Mpc$^{-1}$,
$\Omega_{\Lambda} = 0.73$, and $\Omega_m = 0.27$ is adopted
\citep{kom09}.

\section{Data and Data Analysis}

\subsection{The Sample}

Our sample includes all the BLRGs observed by modern X-ray astronomy
satellites ({\it EXOSAT}, {\it Ginga}, {\it ASCA}, {\it RXTE}, {\it
BeppoSAX}, {\it Chandra}, {\it XMM-Newton}, {\it Suzaku}, {\it INTEGRAL}
and {\it Swift}), for which data are available at energies above
2\,keV. Table\,1 presents the list of 18 BLRGs compiled, which also
includes for comparison a sample of nine bright Seyfert 1 galaxies
chosen from the compilation by \citet{ho01} and \citet{ho02}, 
for which the nuclear optical and hard X-ray fluxes match those
of the discussed BLRGs. Here we have 
selected luminous Seyferts 1 with measured radio fluxes (or meaningful upper limits) which, 
due to their high accretion rates ($>1\%$ Eddington) and unobscured nuclei, 
may be considered as radio-quiet analogues of BLRGs \citep[see the discussion 
in][]{sik07}\footnote{In our sample of Seyfert galaxies we have therefore
excluded type $1.5-2$ objects, in order to avoid additional
complications related to the absorption of the X-ray emission by the
circumnuclear cold gas and dust. We have also excluded sources such as
NGC~4639, which, even though classified as type 1 Seyfert \citep{sik07},
is an example of a low-luminosity AGN accreting at a low rate
\citep{ho99}.}. The main difference between the two analyzed classes of
sources that should be emphasized once more is that Seyferts are
radio-quiet and hosted by late-type galaxies, while BLRGs are very
radio-loud and elliptical-hosted. The other possibly relevant (but
related to the morphologies of host galaxies) difference regards the masses of
their SMBHs, $\mathcal{M}_{\rm BH}$. In particular, nine Seyferts included in
our sample are characterized by lower values of black hole masses
(median $\approx 10^{6.9} M_{\odot}$) when compared to the ten BLRGs
considered here with $\mathcal{M}_{\rm BH}$ provided in the literature
(median $\approx 10^{8.8} M_{\odot}$; see Table\,1).

The basic information collected from the literature regarding each
analyzed source, as listed in Table\,1, are (1) IAU coordinates for
J2000, (2) source name, (3) redshift $z$, (4) luminosity distance
$d_{\rm L}$, (5) total 5\,GHz flux $[\nu F_{\nu}]_{\rm 5\,GHz}^{\rm
tot}$ in the cgs units of erg\,cm$^{-2}$\,s$^{-1}$, (6) 5\,GHz flux of
the unresolved nucleus $[\nu F_{\nu}]_{\rm 5\,GHz}^{\rm nuc}$, (7)
$B$-band optical flux of the nucleus $[\nu F_{\nu}]_{\rm B}^{\rm nuc}$,
(8) X-ray photon index $\Gamma_{\rm X}$ measured between 2 and 10\,keV,
(9) average $2-10$\,keV flux $[\nu F_{\nu}]_{\rm 2-10\,keV}$, (10) hard
X-ray/soft $\gamma$-ray flux $[\nu F_{\nu}]_{\rm 14-195\,keV}$ detected
by {\it Swift}-BAT at $14-195$\,keV, (11) black hole mass
$\mathcal{M}_{\rm BH}$, and (12) references. In the case of
Seyferts the provided nuclear $B$-band fluxes \citep[same as
in][]{sik07}, which are carefully corrected for the non-negligible
starlight contamination, are taken from \citet{ho01} and
\citet{ho02}. In the case of BLRGs, much less severe starlight
contamination was taken into account by means of the appropriate
correction factors given by \citet{era94} and \citet{era03}. Also, when
a $B$-band flux was not provided in the literature explicitly, we
estimated it from a $V$-band flux applying a multiplication factor
($\lambda_{\rm V}/\lambda_{\rm B})^{1 - \alpha_{\rm opt}}$ with
$\lambda_{\rm V} = 5500$\,\AA, $\lambda_{\rm B} = 4400$\,\AA, and the
assumed optical power-law slope $\alpha_{\rm opt} = 0.5$. Finally, all
the $B$-band nuclear fluxes listed in column 7 have been corrected for
the Galactic extinction available in
NED\footnote{\texttt{http://nedwww.ipac.caltech.edu}}. We note that
BLRG 3C~227 has been detected only very recently by {\it Swift}-BAT at a
high significance level \citep[S/N of 4.87;][]{tue10}, but is poorly
known at other wavelength. In X-rays, two {\it Chandra} observations of
this object have been conducted so far, as parts of a survey of multiple
hot spots in the large-scale structures of nearby radio galaxies. We
estimated the $2-10$\,keV nuclear flux of 3C~227 from the count rate of
{\it Chandra} ACIS CCD chip given in \citet{har07} using the software
\textsc{PIMMS}\footnote{\texttt{http://heasarc.nasa.gov/Tools/w3pimm.html}}
and assuming the X-ray photon index $\Gamma_{\rm X} = 1.5$.
\footnote{The uncertainty in the assumed X-ray photon index $\pm 0.5$
would make only $\simeq 20\%$ difference in the estimated $2-10$\,keV flux.}

We are aware that the compiled sample of BLRGs is by no means 
statistically complete or unbiased, since the targets were 
originally selected from independent programs by different observers
using different instruments. The final list of objects reflects in fact a
bias toward sources which are the brightest in the X-ray domain. Moreover, we
are aware that most of the considered objects show flux variability in
various energy bands. For example, \citet{tur89} recorded the minimum
X-ray flux of 3C~111 at the level of $1.8 \times
10^{-11}$\,erg\,cm$^{-2}$\,s$^{-1}$, whereas the maximum flux for this
source recorded a decade later reached $5.6 \times
10^{-11}$\,erg\,cm$^{-2}$\,s$^{-1}$ in the $2-10$\,keV band
\citep{era00}. Table\,1 lists the \emph{average} fluxes measured at
different wavelengths as provided in literature, while at the same time
the results discussed later in this paper are not affected by the
relatively modest (typically up to a factor of 3) flux variations
characterizing X-ray and radio continua of BLRGs\footnote{In order to 
illustrate possible effects caused by the temporal variability of the 
analyzed sources, in the spectral energy distributions shown in Figures\,3, 
8 and 9 below we have provided both the minimum and maximum X-ray 
fluxes whenever available in the literature.}. The same caveats apply to 
the sample of Seyfert galaxies analyzed as it was compiled using even 
more heterogeneous criteria. Hence it is difficult to identify and
to discuss all the possible biases introduced by the applied selection 
criteria for both samples. Therefore these should be considered simply as
lists of prominent (bright) examples of the two discussed classes of AGN.

\begin{figure}
\begin{center}
\includegraphics[angle=0,scale=0.47]{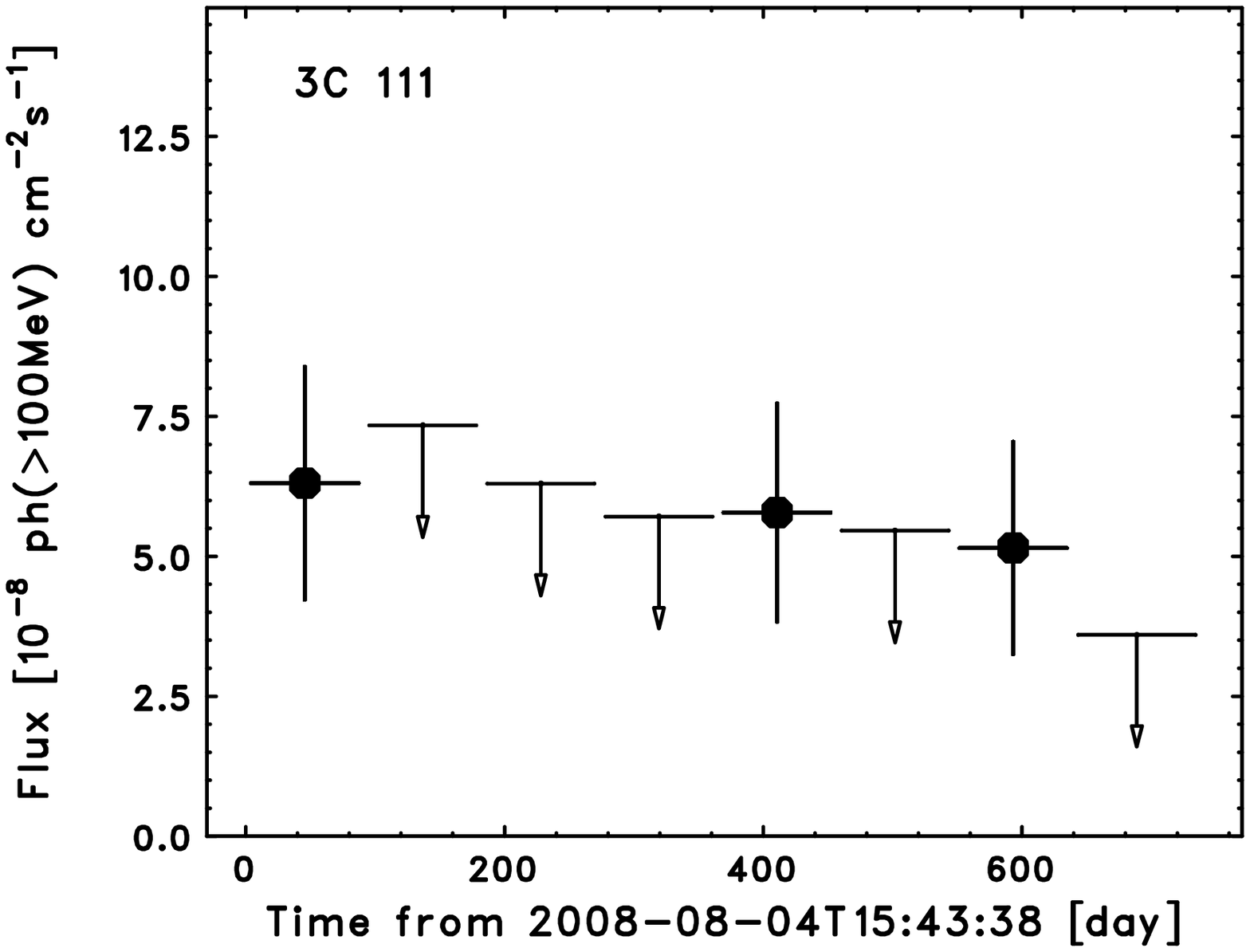}
\includegraphics[angle=0,scale=0.47]{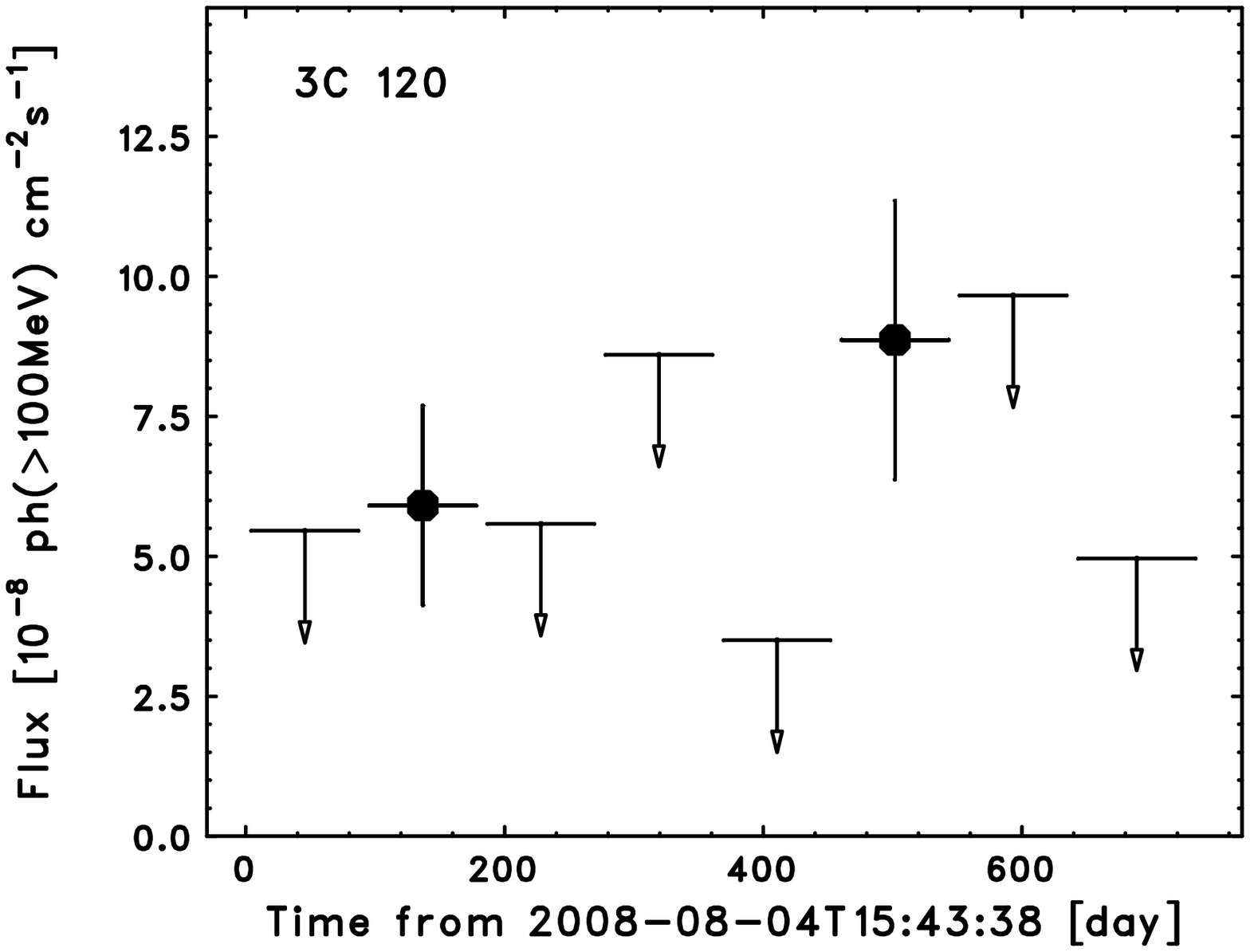}
\caption{Temporal variation of $\gamma$-ray flux ($>100$\,MeV) of 3C~111
 ({\it top}) and 3C~120 ({\it bottom}) over the period 2008 August--2010
 August. The time (in days) is measured from the start of the \F
 observation (2008 August 4, 15:43:38 UT). The fluxes are plotted
 only when the corresponding TS values exceed 10 for a given time bin,
 otherwise upper limits are provided.}
\end{center}
\end{figure}

\subsection{\F Observations and Data Analysis} 

The LAT instrument onboard {\it Fermi} is described in detail in
\citet{atw09} and references therein. It is characterized by a larger
effective area ($\sim$8,000\,cm$^2$ on axis at $1$\,GeV for the event
class considered here), a wider energy coverage (from $\sim$20\,MeV to
$>$300\,GeV), and an improved angular resolution when compared to the
previous $\gamma$-ray missions. The $68\%$ containment angles of the
reconstructed incoming photon direction are approximated as
$\theta_{68}\simeq0^{\circ}.8\,(\varepsilon_{\gamma}/{\rm GeV})^{-0.8}$
below $10$\,GeV. During the first two years of \F operation, most of the
telescope's time was dedicated to observing in a `survey mode,' in which
the instrument points away from the Earth and nominally rocks the
spacecraft axis north and south from the orbital plane to enable
monitoring of the entire sky on a time scale shorter than a day. In
particular, the whole sky is surveyed every $\sim 3$\,hours (2
orbits).

The dataset used here comprises all the scientific data obtained between
August 4, 2008 and August 4, 2010. This time interval runs from Mission
Elapsed Time (MET) 239557414 to 302630530, which is consistent with the
observation period for the Second \F Catalog (2FGL) selection (The 
Fermi-LAT collaboration 2011, in prep). We have applied the zenith angle cut of
$105^{\circ}$ to eliminate photons from the Earth's limb. 
We use events from the ``Diffuse'' class \citep{atw09}, i.e., 
the events that have the high probability of being
photons. In the analysis presented here, we set the lower and higher
energy bounds at $200$\,MeV and $100$\,GeV, respectively. The choice of
a lower energy bound at 200 MeV is conservative but reduces
significantly systematic errors. Science Tools\footnote{\texttt{http://fermi.gsfc.nasa.gov/ssc/data/analysis/documentation/Cicerone/}} version \textsc{v9r15p2}
and Instrumental Response Functions (IRFs) \textsc{P6\_V3\_DIFFUSE} were
used throughout the analysis.

In order to study the average spectrum of each selected target, we use
the standard unbinned 
maximum-likelihood spectral estimator \citep{mat96} provided with the LAT
science tool \textsc{gtlike}. It allows us to fit the data to a source
model, along with the models for the uniform extragalactic and
structured Galactic
backgrounds\footnote{\texttt{http://fermi.gsfc.nasa.gov/ssc/data/access/lat/BackgroundModels.html}}.
The Galactic diffuse emission model and the isotropic spectral
model used here were developed by the \F team as refinements of the
publicly released models; this choice does not affect significantly the
results for the candidate sources considered, which are all located
outside the Galactic plane except for 
3C~111\footnote{With Galactic coordinates ($l = 161^{\circ}.68$, $b = -8^{\circ}.82$), 
3C~111 is located behind a large molecular cloud complex 
in Taurus \citep[cf.,][Fig.~2 therein]{dam01}, 
thus requires more care in the analysis due to the expected contamination from Galactic diffuse emission.\label{footnote:3c111}}. 
Much more crucial is a careful selection of 
source regions, especially for relatively faint objects. The model for
which we calculate the likelihood is a combination of point-like and
diffuse sources for a region of interest (ROI) with the radius of $r =
8^{\circ}$ centered on the target under consideration\footnote{See, in
this context, \citet{PerA} for a detailed investigation of changing the
radius of ROI from $8^{\circ}$ to $20^{\circ}$, and the arguments given
for using $r=8^{\circ}$ to minimize the contamination from the Galactic
diffuse emission.}. For all the BLRGs and Seyferts listed in
Table\, 1, we first assume a point-like source with a power-law spectrum
at the position of each target, and fix the $\gamma$-ray photon index as
$\Gamma_{\gamma} = 2.5$. Additional sources from the 1st \F Catalog
\citep[1FGL;][]{1FGL} and from the internal LAT collaboration catalog
produced using 18 months 
of data are included in the model of each ROI. Next, using the
\textsc{gtlike} tool, we find the best-fit parameters for each source
and evaluate the significance of the detection given by the test
statistic, ${\rm TS} = 2 \Delta \log({\rm likelihood})$ between the
models with and without a source. For sources which are detected 
above a certain significance threshold 
(${\rm TS} \geq 25$, corresponding to $\geq 5\sigma$
detections), we retry the \textsc{gtlike} fit assuming power-law spectra
($F = K \, E^{- \Gamma_{\gamma}}$) with \emph{both} parameters
(normalization, $K$ and photon index, $\Gamma_{\gamma}$) set free, and
then calculate the errors on the fluxes and photon indices. For the
remaining BLRGs and Seyferts, which are detected in the analyzed dataset
below the threshold (${\rm TS} < 25$), we simply
provide upper limits on the fluxes for the fixed $\Gamma_{\gamma} =
2.5$.

\begin{figure}
\begin{center}
\includegraphics[angle=0,scale=0.22]{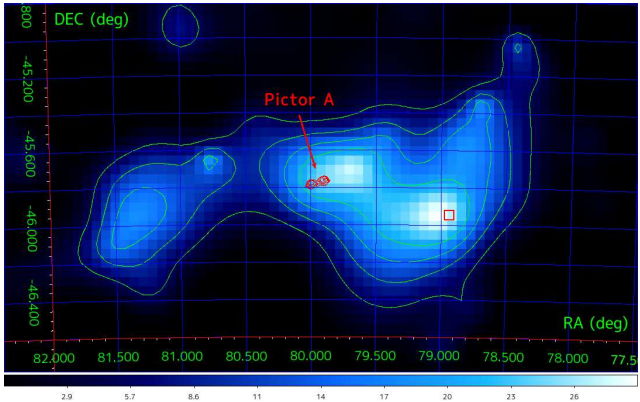}
\caption{
\F TS map ($>$200 MeV) centered on Pictor A, showing the presence
of multiple $\gamma$-ray peaks in the field. Contamination from nearby 
sources are modeled as part of the background that also includes 
Galactic/extragalactic diffuse $\gamma$-ray emission.
The $green$ contours correspond to TS values of 5, 10, 15, and 20, 
while the $red$ contours indicate 1.4 GHz radio emission from 
Pictor A \citep{per97}. The position of a blazar, BZQJ0515-4556, is also
 shown as $red$ $box$. The peak near the center of the map (TS = 20) 
is almost exactly coincident with the position of Pictor A. 
The TS peaks associated with Pictor A and with BZQ J0515$-$4556 are 
only marginally resolved.  The TS value quoted here and in Table 2 
for Pictor A was evaluated using a source model that included 
point sources at the position of BZQ J0515$-$4556 and at 
R.A. = 81.35 deg, Dec. = $-$45.78 deg.
}
\end{center}
\end{figure}

\section{Results}

Table\,2 summarizes the results of the \F data analysis of the 18
BLRGs and 9 Seyfert galaxies. For each source considered, Table\,2
provides (1) name, (2) statistical significance ${\rm TS}$ of the \F
detection, (3) $\gamma$-ray photon index $\Gamma_{\gamma}$ evaluated for
the photon energy range $0.1-10$\,GeV, (4) the integrated photon flux
above $100$\,MeV, $F_{\rm >0.1\,GeV}$, (5) $\gamma$-ray flux $[\nu
F_{\nu}]_{\rm 0.1-10\,GeV}$,  (6) the corresponding $\gamma$-ray
luminosity $L_{\gamma} = 4 \pi d_{\rm L}^2 [\nu F_{\nu}]_{\rm
0.1-10\,GeV}$,  (7) total accretion-related luminosity $L_{\rm acc}$
derived from the spectral fitting as described below, 
and (8) `mixing' parameter $\eta$ discussed in the next section.
Only two BLRGs (3C~111 and 3C~120) are detected at
sufficiently high significance levels, i.e., ${\rm TS} \geq 25$, in the
accumulated two-year \F dataset. For these, the $\gamma$-ray fluxes and
luminosities are evaluated straightforwardly\footnote{Since the
likelihood analysis was limited to the photon energy range
$0.2-100$\,GeV, all the flux values and the corresponding luminosities
are extrapolated down to $100$\,MeV with a given photon index. This
choice is dictated solely by the \emph{convention} typically followed by
the $\gamma$-ray community.}. For the targets detected at lower
significance levels, ${\rm TS} < 25$, the corresponding $95\%$
confidence level flux upper limits are calculated using the dedicated
software \textsc{UpperLimit.py}.

\begin{figure}
\begin{center}
\includegraphics[angle=0,scale=0.47]{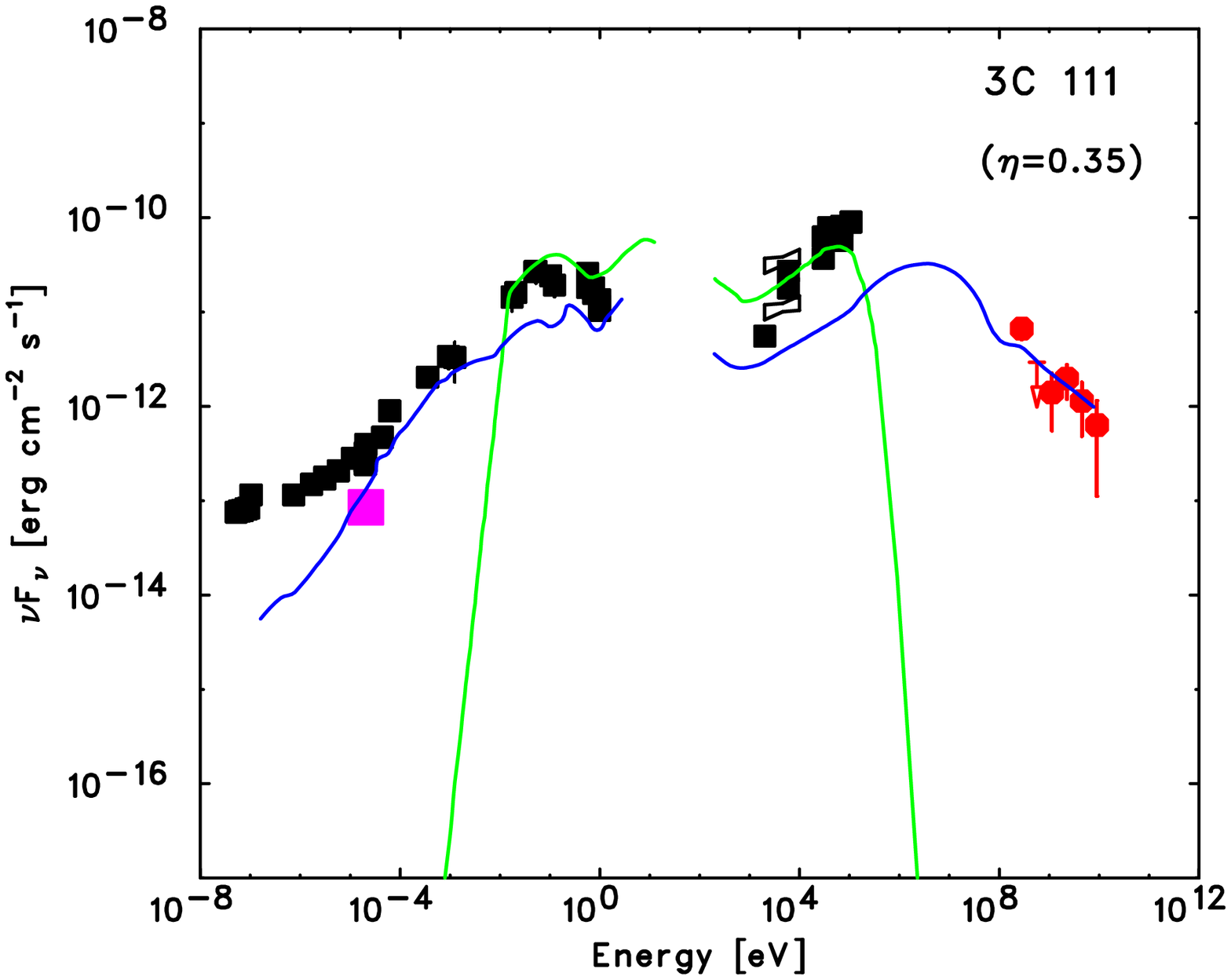}
\includegraphics[angle=0,scale=0.47]{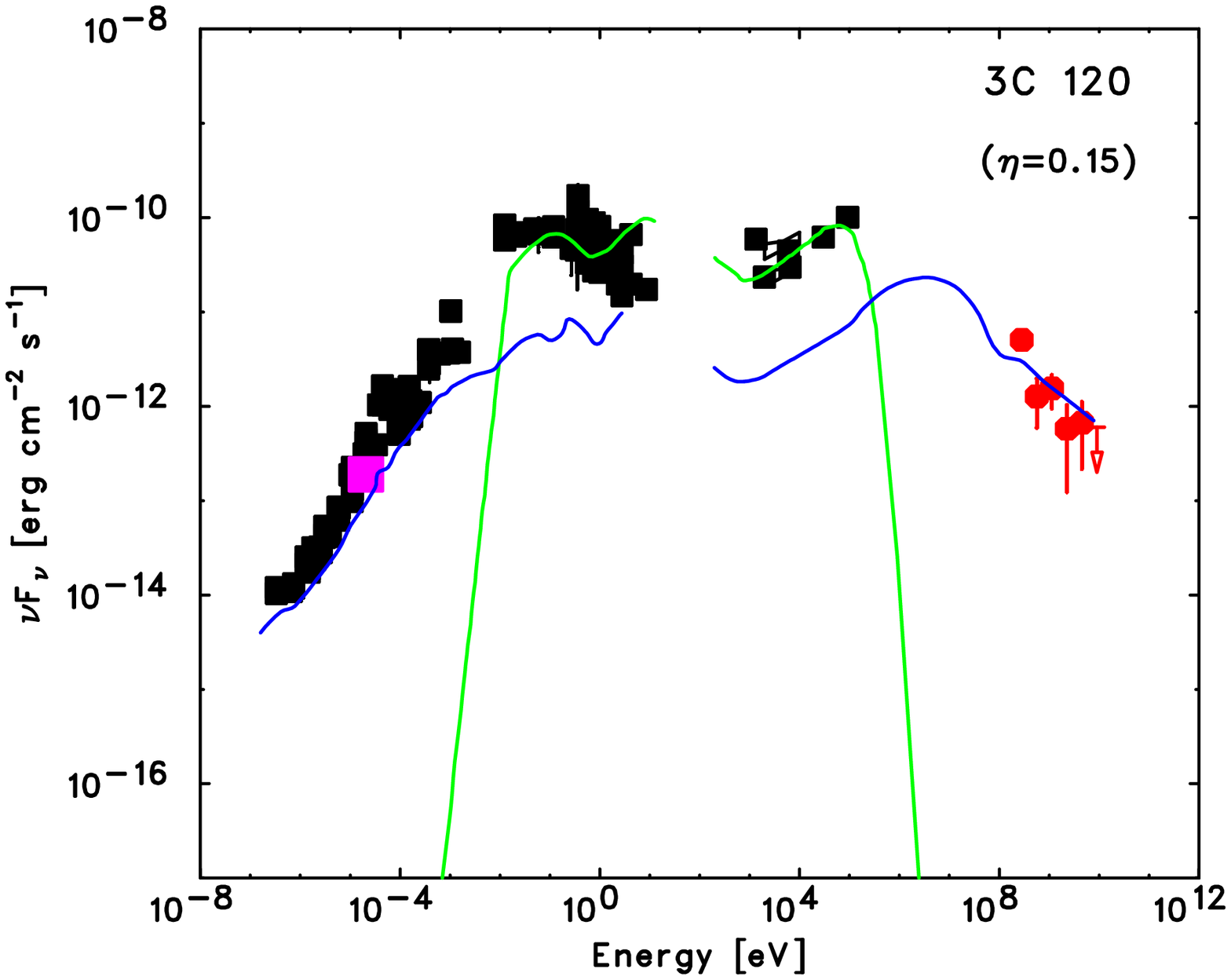}
\caption{Broad-band SEDs of the two BLRGs detected at high significance by \F (3C~111 and 3C~120). \F data are indicated by red circles. Black squares represent the historical data from NED. Magenta squares denote the $5$\,GHz radio fluxes of the unresolved nuclei (if available). 
The green curves correspond to the template of the accretion-related
 Seyfert-type emission \citep[from][]{kor99}, matched to the
 infrared--to--X-ray continuum of each source. The blue curves correspond
 to the broad-band spectrum of the quasar 3C~273 \citep[from][]{sol08},
 used here as a template of the jet-related emission and scaled to match
 the radio fluxes for each source. The mixing parameter $\eta$ for the
 phenomenological hybrid model discussed in \S\,4 is given in each
 panel.}
\end{center}
\end{figure}

\begin{figure}
\begin{center}
\includegraphics[angle=0,scale=0.47]{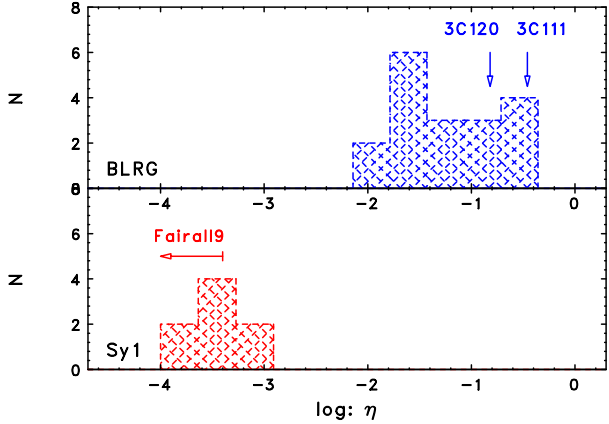}
\caption{
Distribution of a mixing parameter $\eta$ for BLRGs ({\it top} panel;
 blue histogram) and Seyferts ({\it bottom} panel; red histogram).
The values characterizing the two detected sources (3C~111 and 3C~120) 
are indicated by the arrows in the top panel. In the case of Fairall~9,
 which is not detected in radio, an upper limit for the $\eta$ parameter
 is given in the {\it bottom} panel.}
\end{center}
\end{figure}

For 3C~120, the results presented here are consistent with those
reported in \citet{MAGN}, but with reduced uncertainties in the flux and
photon index due to the improved photon statistics
based on the two-year accumulation of the \F data. We note however
that the ${\rm TS}$ value increased only slightly between the 15-month
and 24-month \F datasets \cite[cf., ${\rm TS} = 34$ found here
vs. ${\rm TS} = 32$ reported in][]{MAGN}. 
In contrast, the flux and photon index uncertainties 
increased in the case of 3C~111, and the corresponding $
{\rm TS}$ value decreased between the 15-month and 
24-month \F datasets \cite[cf. ${\rm TS} = 31$ found here vs. ${\rm
TS} = 34$ reported in][]{MAGN}. The reason for this behavior is 
twofold. First, the likelihood analysis was limited here to the photon
energy range $0.2-100$\,GeV, whereas the energy range $0.1-100$\,GeV was
adopted in \citet{MAGN}. The difference in energy selection is relevant
since 3C~111 is located at a relatively low Galactic latitude (see footnote~\ref{footnote:3c111}), and as such is heavily affected by the contamination from
the Galactic diffuse emission especially below $200$\,MeV
\footnote{Accordingly, in the 1FGL catalog \citep[][table~4 therein]{1FGL}, 
the source fit of 1FGL~J0419.0+3811 was flagged as being sensitive to 
changes in the diffuse Galactic emission model (flux and spectral index could change by more than 3$\sigma$). 
However, upon close inspection of the dust column density 
$E(B-V)$ and $W$(CO) maps \citep{sch98,dam01}, 
we estimated that the possible enhancement of $\gamma$-ray emission at the 
position of 3C~111 due to additional column density from dark gas 
or a somewhat larger emissivity may account for at most, $\sim$10 $\%$ of
the $\gamma$-ray flux listed in Table~2.}.
If we \emph{lower} the photon energy threshold of the likelihood analysis to
$100$\,MeV, the significance of the 3C~111 detection in the 24-month data increases
(${\rm TS} = 59$), as expected. 
Second, as noted in \citet{MAGN}, the considered
galaxy seemed variable in the GeV range, being in particular brighter at
the very beginning of \F observations. 
Here we investigate this issue in more detail, showing in Figure\,1
({\it top} panel) the temporal variations of the $\gamma$-ray 
flux of 3C~111 above 100\,MeV using the two-year accumulation of 
the \F data binned in the 3-month-long periods. 
The fluxes are plotted only when the detection significance
reached ${\rm TS} > 10$ in a given time bin; in the case of 3C~111 such
a criterion was fulfilled only in three time intervals out of 8
total. Even lower $\gamma$-ray duty cycle emerges from the 3C~120
lightcurve, as also shown in Figure\,1 ({\it bottom} panel). All in all,
we conclude that both BLRGs detected by \F are variable in the GeV
photon energy range, even though the significance of the variability can
hardly be evaluated due to the insufficient photon statistics.
A dedicated analysis of all EGRET data also indicated plausible 
variability of the MeV/GeV emission in 3C~111 \citep{har08}. 
Note also that over the past two years both BLRGs have declined in flux by 
$\sim$30\% at centimeter wavelengths as observed by the Univ. of
Michigan Radio Astronomy
Observatory\footnote{http://www.astro.lsa.umich.edu/obs/radiotel/umrao.php}.
In 3C~111 specifically, a bright sub-mm (230 GHz) flare was observed 
toward the end of 2008 \citep{cha11}, 
coinciding with the period of the initial $\sim$0.5 years of LAT
observations, with a subsequent decline in the sub-mm flux over the
next $\sim$1.5 years.

At this point let us also comment on the particular case of
Pictor~A galaxy, for which a relatively high TS value has been found
(see Table\,2), yet below the critical value of 25 required for claiming
the detection. In fact, in the analysis procedure described above (i.e.,
assuming a \emph{single} point-like source at the position of the
galaxy, ${\rm R.A.} = 79^{\circ}.957$ and ${\rm Dec} =
-45^{\circ}.779$), one can find formally ${\rm TS} = 45$ for
Pictor~A. However, a close inspection of the spatial distribution of TS
for the whole source region (hereafter the `TS map') reveals a rather
complex structure elongated substantially in the RA direction,
which is inconsistent with a single point source. Figure\,2 shows 
TS map thus obtained, where contamination from nearby sources
(listed in the 1FGL and in the internal LAT collaboration catalog
using 18 months of data) are modeled as part of the 
background that also includes 
Galactic/extragalactic diffuse $\gamma$-ray emission. This suggests a 
presence of multiple $\gamma$-ray sources in the field. More exactly, in
a careful examination of the TS map we found three emission
peaks (each about a degree apart), one of which coincides almost exactly
with the position of Pictor~A, being characterized by ${\rm TS} =
20$. Also, one of the other two emission peaks located at ${\rm R.A.} =
79^{\circ}.09$, ${\rm Dec} = -46^{\circ}.08$ is positionally
coincident with a blazar BZQ~J0515$-$4556 (${\rm R.A.} = 78^{\circ}.94$, 
${\rm Dec} = -45^{\circ}.95$) with a TS value of 19. The last peak 
seen to the east of Pictor A is less prominent (TS value of 11) 
and located at ${\rm R.A.} = 81^{\circ}.35$, ${\rm Dec} =
-45^{\circ}.78$. Sources at the positions of the two TS peaks 
just mentioned were included in the model when the TS for Pictor A 
listed in Table 2 was evaluated.
Hence, one may conclude that there are some indications for the GeV emission of
Pictor~A galaxy close to the upper limits provided in Table\,2, and that
the formal detection of this object by \F in the near future is quite
likely.

\begin{figure}
\begin{center}
\includegraphics[angle=0,scale=0.47]{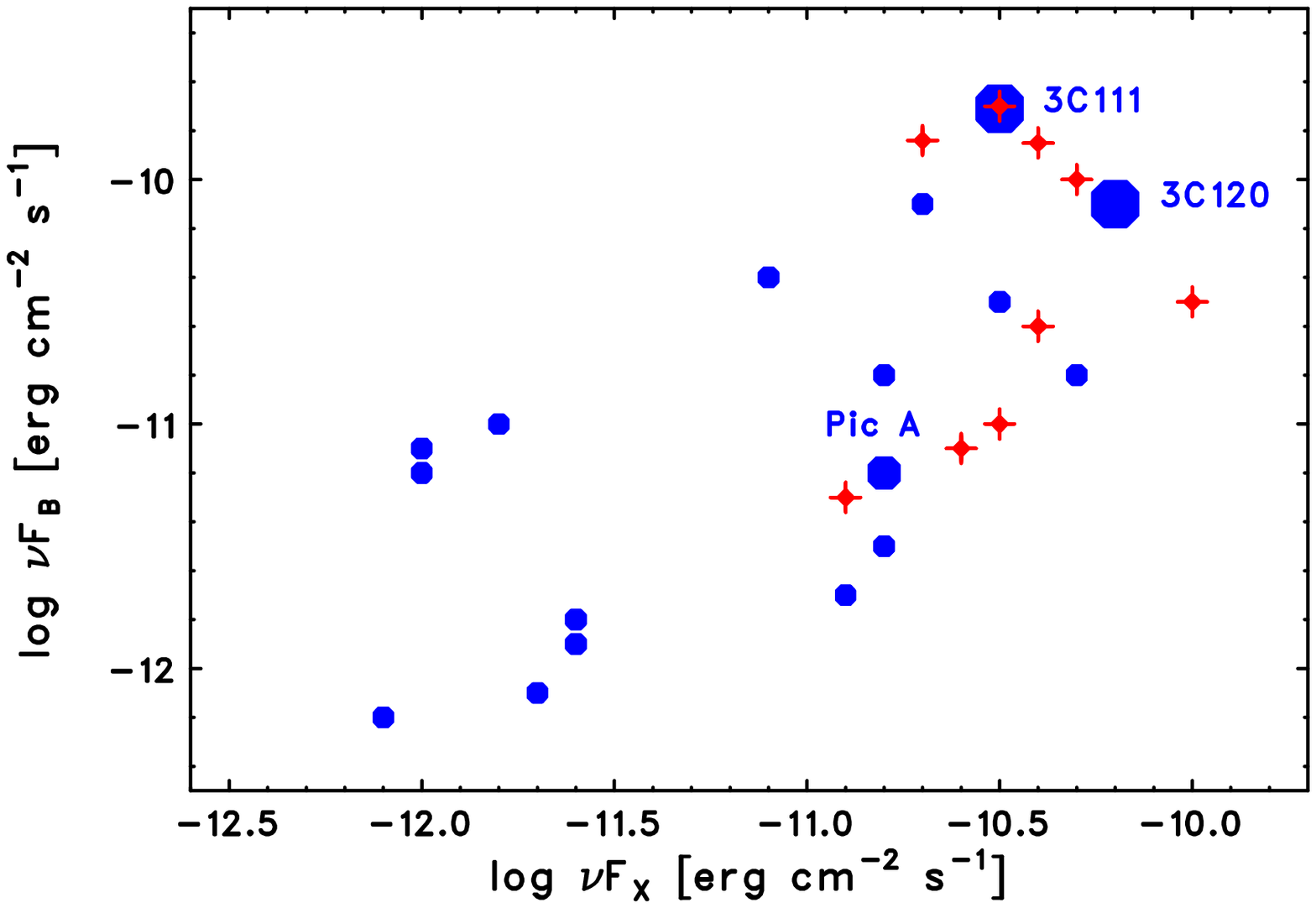}
\includegraphics[angle=0,scale=0.47]{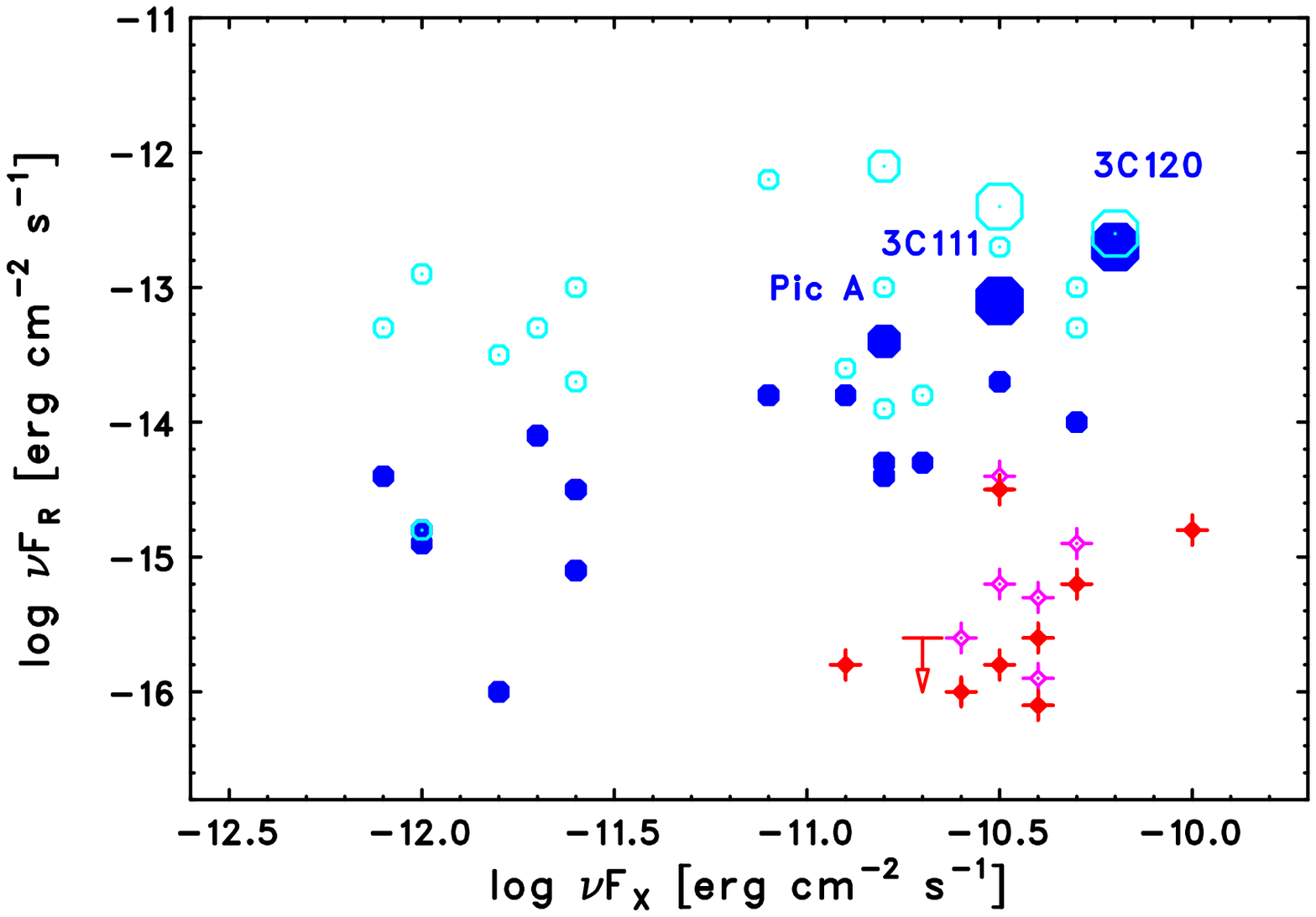}
\caption{{\it Top} panel: the optical
 ($B$-band) versus X-ray ($2-10$\,keV) fluxes 
for all the sources included in the sample, with BLRGs
 denoted by blue filled circles and Seyferts by red crosses. {\it
 Bottom} panel: the radio ($5$\,GHz) versus X-ray ($2-10$\,keV)
 fluxes. Here BLRGs are plotted as blue filled cycles when nuclear 
radio fluxes are used, and as cyan open circles when total radio fluxes are
 considered. Similarly, Seyfert 1 galaxies are plotted as red crosses
 when nuclear radio fluxes are used, and as magenta crosses when total
 radio fluxes are considered. An upper limit is given for Fairall~9. 
In both panels, large symbols indicate the two objects
 detected by \F (3C~111 and 3C~120), while the medium-size symbols
 represent Pictor A galaxy which, even though not formally detected, is
 characterized by the third-highest TS in the analyzed two-year \F
 dataset.}
\end{center}
\end{figure}

No statistically significant detection of any Seyfert 1 galaxy analyzed
was found in the 24-month \F dataset. Interestingly, IC~4329A shows a
relatively high ${\rm TS}$ value of $15$ (see Table\,2), but this
could be due to a contamination from a nearby source. This nearby 
source was found only recently, and has been associated tentatively with
a blazar 
in the 2FGL catalog (The Fermi-LAT collaboration 2011, in
preparation). Nevertheless, the provided upper limits for the
$\gamma$-ray emission of the considered sample of Seyferts --- typically
at the flux level of a few $\times 10^{-12}$\,erg\,cm$^{-2}$\,s$^{-1}$
between 0.1 and 10\,GeV, depending on the degree of the contamination
from the Galactic diffuse emission and on the presence of nearby bright
$\gamma$-ray sources --- provide interesting constraints as discussed
below.

\begin{figure}
\begin{center}
\includegraphics[angle=0,scale=0.47]{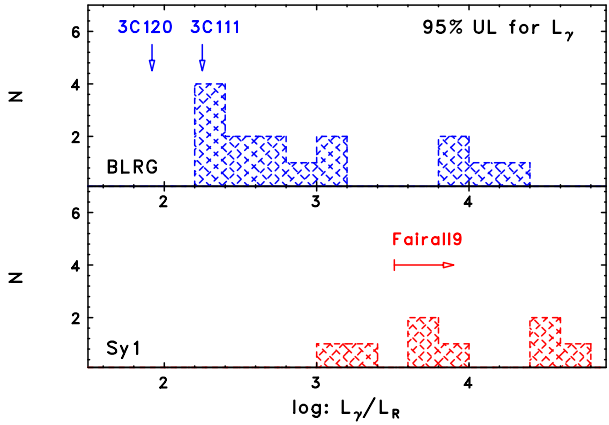}
\caption{Upper limits for the ratio of the monochromatic
 $\gamma$-ray and nuclear radio luminosities for BLRGs ({\it top} panel;
 blue histogram) and Seyferts ({\it bottom} panel; red histogram)
 that are not detected by \F. The values characterizing the two detected sources
 (3C~111 and 3C~120) are indicated by the arrows in the top panel. 
Since Fairall~9 is not detected in radio, a lower limit 
is given in the {\it bottom} panel.}
\end{center}
\end{figure}

\section{Discussion and Conclusions}

The analysis of the two year \F data for the selected 18
X-ray--bright BLRGs confirmed the previously reported detections of
the two sources (3C~120 and 3C~111), at the significance levels not much
different between the 15-month and 24-month datasets. This may indicate 
that the observed $\gamma$-ray emission is variable on months to year
timescales, with a relatively low duty cycle, as revealed by a closer
inspection of the corresponding lightcurves (Figure\,1).  
In fact, the $>100$ MeV $\gamma$-ray flux of 3C~111 observed with \F is 
(3.5$\pm$1.2)$\times$10$^{-8}$ ph cm$^{-2}$
s$^{-1}$, which is about $\sim 20\times$ smaller than the maximum 
recorded by EGRET in the same energy range 
\citep[64$\times$10$^{-8}$ ph cm$^{-2}$ s$^{-1}$,][]{har08}, thus 
suggesting significant variability in the decade timescale between the EGRET and \F observations.
Moreover, we found some hints for $\gamma$-ray emission 
from Pictor A (${\rm TS} = 20$), 
even though we cannot claim a formal detection at the moment.
These results suggest that other BLRGs could be promising candidates 
for \F detections in the near future, and that BLRGs in general 
could potentially constitute a class of $\gamma$-ray loud AGN. 
Such a statement is however difficult to
quantify, because of the aforementioned variability of the GeV fluxes
from 3C~120 and 3C~111 (assuming that both galaxies are representative
for the whole class). On the other hand, the observed flux changes from
these two detected galaxies imply that, in analogy with blazar
sources, the GeV continua of BLRGs are produced predominantly in the
inner parts of their nuclear outflows (jets on scales less than tens or
hundreds of parsecs).  In fact, as noted in the previous
sections, several authors have previously suggested that the GeV
radiation from BLRGs, if detected, should most likely be due to
non-thermal and beamed blazar-type emission, but only observed at
intermediate inclinations \citep[jet viewing angles, say, $10^{\circ}
\lesssim \theta_j \lesssim 30^{\circ}$, versus $\theta_j \lesssim
10^{\circ}$ expected for blazars; see][]{gra07,sam09}. 
Interestingly, unlike in the luminous blazars, the X-ray/soft
$\gamma$-ray spectra of BLRGs (up to the observed photon energies of
$\sim 100$\,keV) have been argued to be dominated by the thermal
radiation of the accreting matter \citep{woz98,sam99,era00,zdz01}, with
only little (if any) jet contribution \citep{gra07,kat07,sam09}. Hence
it is clear that BLRGs are truly ideal targets for investigating the AGN
jet-disk connection in the X-ray/$\gamma$-ray regime. Here we present a
few considerations regarding this issue.

\begin{figure}
\begin{center}
\includegraphics[angle=0,scale=0.47]{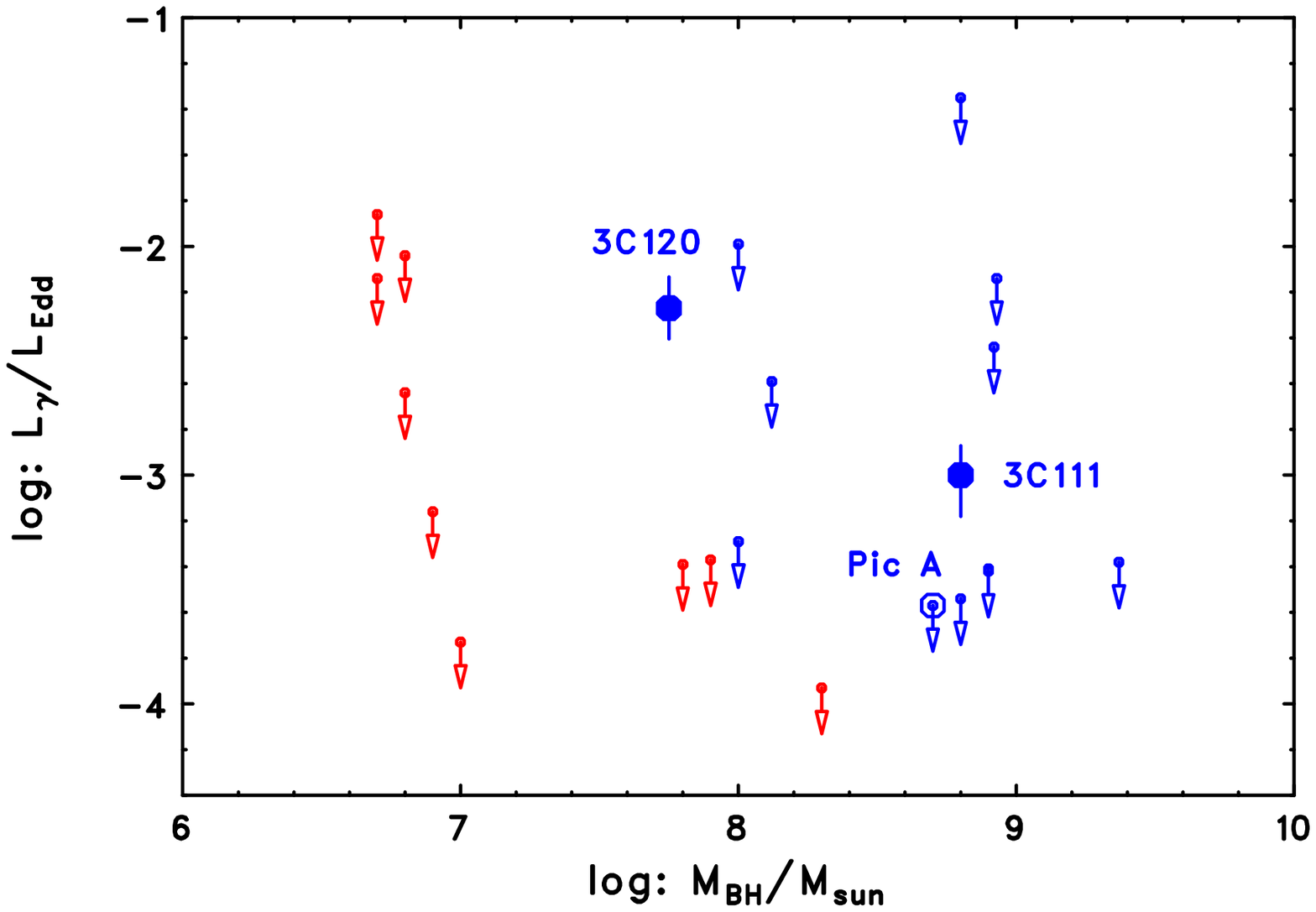}
\includegraphics[angle=0,scale=0.47]{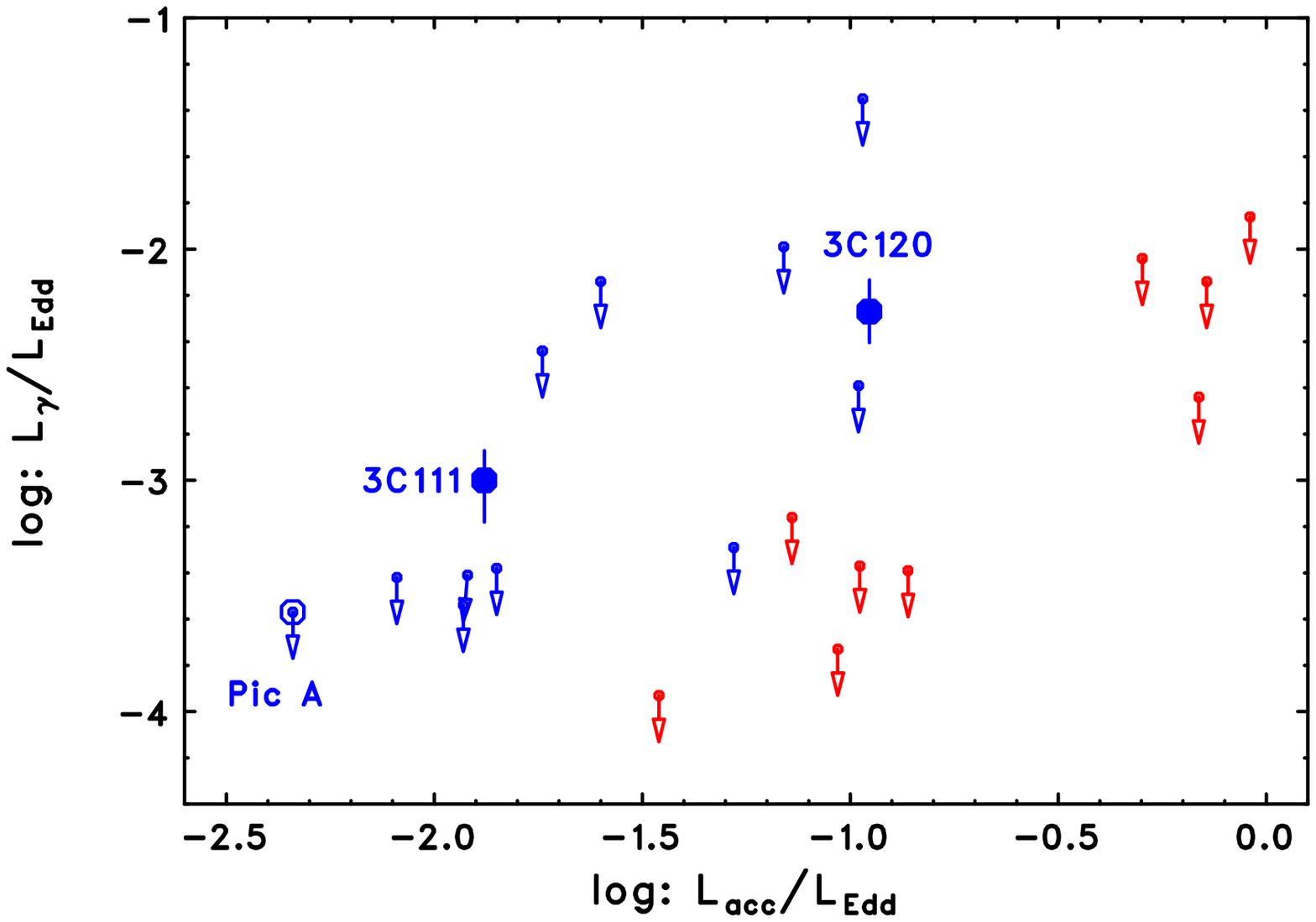}
\includegraphics[angle=0,scale=0.47]{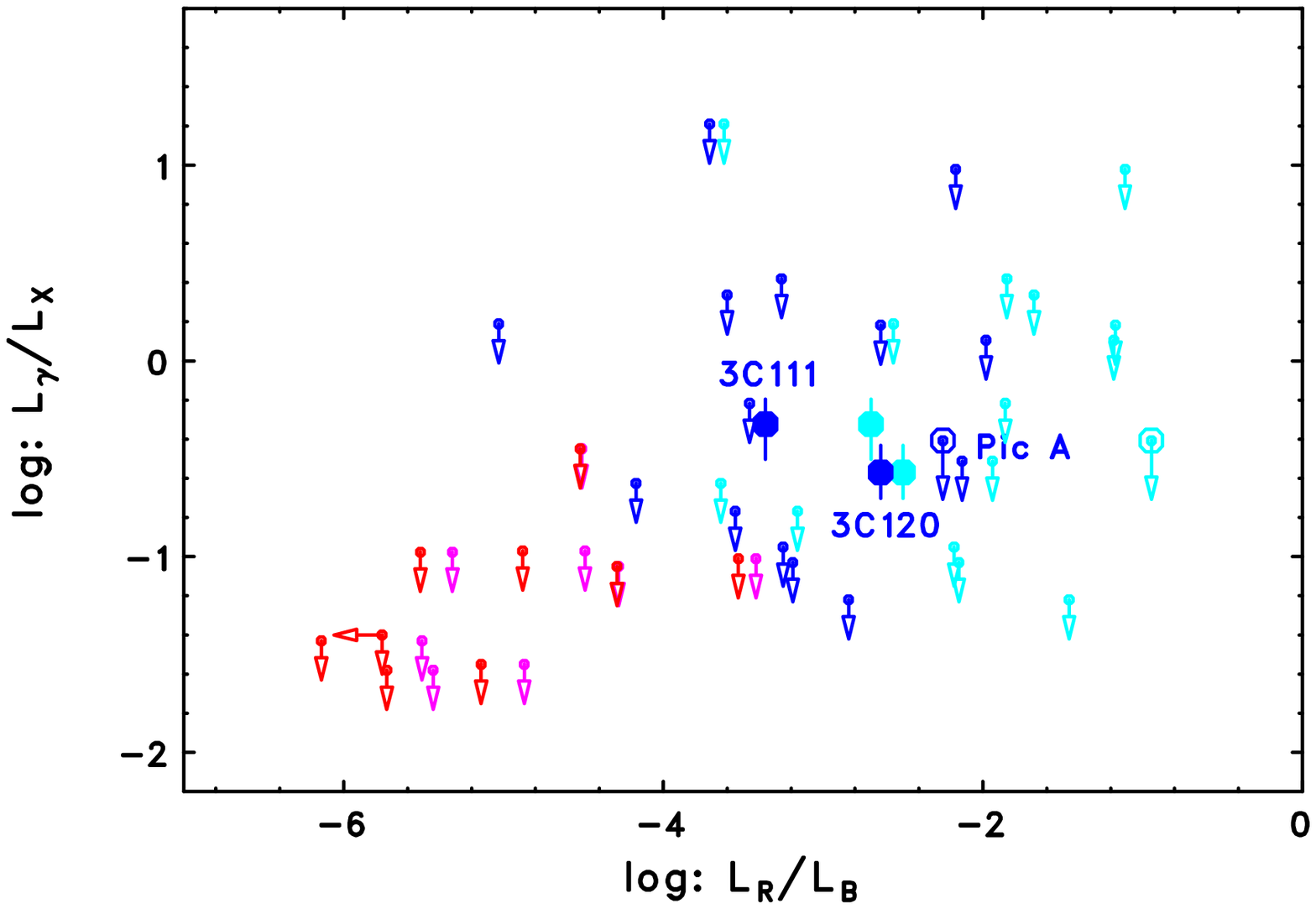}
\includegraphics[angle=0,scale=0.47]{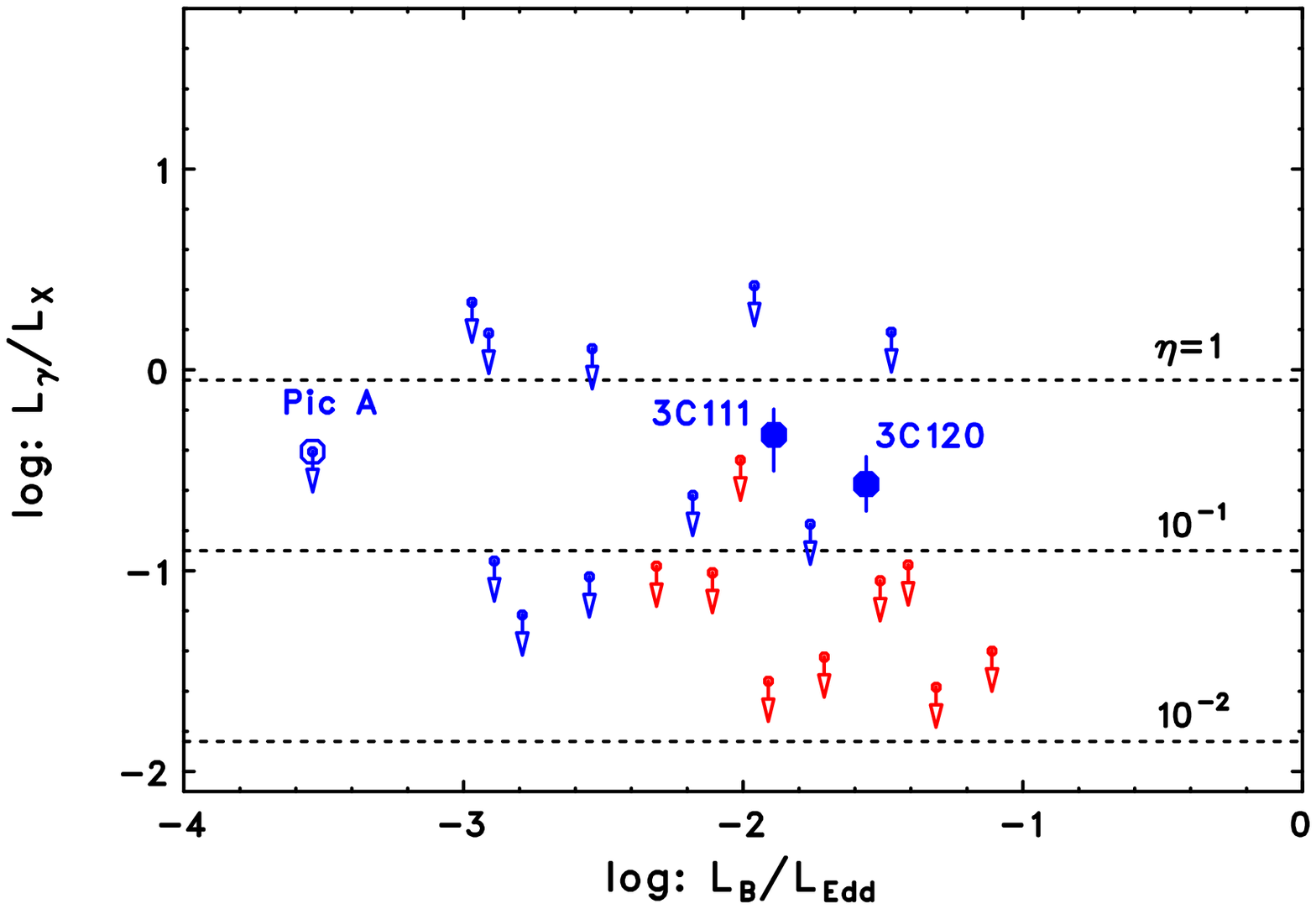}
\caption{The dependence of the $\gamma$-ray luminosities (or upper
 limits for such), expressed in the Eddington units ($L_{\gamma} /
 L_{\rm Edd}$) or as a ratio of the GeV and X-ray monochromatic
 luminosities ($L_{\gamma} / L_{\rm X}$), on 
(i) the black hole mass $\mathcal{M}_{\rm BH}$ ({\it top left} panel),
(ii) the accretion ratio $L_{\rm acc}/L_{\rm Edd}$ estimated from the SED 
fitting ({\it top right} panel), 
(iii) the proxy of the radio loudness parameter $L_{\rm R} / L_{\rm B}$ 
({\it bottom left} panel),  and (iv) the observed proxy of the accretion rate 
$L_{\rm B} / L_{\rm Edd}$ ({\it bottom right} panel). 
BLRGs and Seyferts are denoted by blue and red symbols,
 respectively. Large filled circles represent the two BLRGs detected by
 \F (3C~111 and 3C~120), while the medium-size open circles represent
 Pictor A. In the {\it bottom left} panel, when total radio fluxes 
are used instead of the nuclear radio fluxes, BLRGs are denoted by 
cyan symbols. Similarly, Seyfert 1 galaxies are denoted by magenta 
symbols when total radio fluxes are used instead of the nuclear radio 
fluxes. An upper limit is given for Fairall 9.
The ratios of the GeV and X-ray monochromatic luminosities 
($L_{\gamma} / L_{\rm X}$) expected from a hybrid model are indicated as
 dashed lines in the bottom panel, for different values of the mixing 
parameter $\eta$ = 10$^{-2}$, 10$^{-1}$, and 1.}
\end{center}
\end{figure}

Figure\,3 presents spectral energy distributions (SEDs) of the two
aforementioned $\gamma$-ray--detected BLRGs; the SEDs of all the 
remaining objects, for which only the upper limits in the \F range 
were derived, are given in Figures\,8 and 9 further below (see 
Appendix A). In the figures, the \F data are indicated by red 
circles (or arrows), black squares represent the historical data from NED, 
while the magenta squares show the $5$\,GHz radio fluxes of the unresolved 
nuclei. Even though the broad-band spectra vary to some degree 
between 3C~120 and 3C~111 (e.g., with respect to the ratio between the
total and the nuclear radio luminosities), one can gauge the main
similarities and differences between the BLRG-type and Seyfert-type  
SEDs. In particular, while the infrared--to--hard X-ray continua of the
two BLRGs are very similar to the one characterizing Seyfert galaxies 
considered here,  the former 
objects are significantly brighter than the representative Seyfert in
the radio and $\gamma$-ray domains. Such a dramatic
difference in the radio loudness is related to the presence of a
relativistic jet, as discussed above, and here we suggest that same is
true regarding the $\gamma$-ray loudness. To investigate it further
more quantitatively (yet still illustrative), we consider a simple
phenomenological `hybrid' model for the broad-band emission of a type 1 AGN,
consisting of individual thermal and the non-thermal emission components
\citep[the approach analogous to the one adopted in][]{gra04,gra07}. The
thermal component, related to the radiative output of the accreting and
circumnuclear matter (accretion disk, disk corona, dusty torus), is
approximated here by the template Seyfert spectrum given by
\citet{kor99}. For the non-thermal broad-band emission component of the
nuclear (blazar-type) relativistic jet observed at intermediate viewing
angles, we use the well-constrained SED of the radio-loud quasar 3C~273
\citep[from][]{sol08} as a template\footnote{It should be emphasized
here that the observed multiwavelength emission of 3C~273 is \emph{not
purely non-thermal} in origin. For example, \citet{gra04} estimated that
about a quarter to half of the $2-10$\,keV flux of this source could be
due to the accreting matter, and not due to the jet. However,
the ``big blue bump'' is clearly visible in the overall SED of the quasar
around UV/soft X-ray frequencies is almost certainly due to the
accretion disk. Nevertheless, the broad-band spectrum of 3C~273 is the
best sampled (and best understood) spectrum of a quasar dominated in
radio and in $\gamma$ rays by the moderately-beamed emission of a
powerful jet observed at intermediate viewing angles \citep[$\theta_j
\sim 10^{\circ}$; see][and references therein]{cou98}. As such, it can
indeed be consider as the best available template for the jet-related
emission of BLRGs.}. Both templates, plotted in Figure\,3 (as well as 
in Figures \,8 and 9) as green
and blue curves, respectively, are rescaled to match the data points
for the analyzed sources. 
If the \F did not detect the source and only upper limits on 
$\gamma$-ray flux were derived, its non-thermal component is 
entirely determined by the 5 GHz nuclear flux (in this context, 
see Table~3 of \citet{ghi05}, who predicted $\gamma$-ray fluxes of 3CR 
FR I radio galaxies from 5 GHz core fluxes).
This allows us to approximate the
relative contribution of the non-thermal and thermal emission components
to the observed SED of each object, in terms of a `mixing'
parameter ($\eta$). Under the adopted definition, this parameter increases
with the increasing contribution of the jet-like component, and equals
unity when both the 3C~273 and the Seyfert template spectra provide
comparable contributions to the observed UV flux of a source. Note that
the direct radiation of the standard (Shakura-Sunyaev) AGN accretion
disk is expected to peak in the UV regime (photon energies $\sim
10$\,eV).

Obviously, the adopted model is an oversimplification of a realistic
situation, as it ignores several complications which may be potentially
relevant. For example, it is at some level questionable to use the
broad-band spectrum of the radio-loud quasar 3C~273 as a template for
the jet-related emission of Seyfert galaxies in general. Moreover,
possible (or even likely) temporal variability of each source
considered, as well as the {\it ad hoc} adopted procedure of matching
the assumed templates to the collected data points, precludes any robust
statistical analysis (e.g., $\chi^2$ fitting) which would allow for
the values and errors of the $\eta$ parameter to be determined more
accurately. Nevertheless, it is worth noting that this simple model
seems to match the SEDs of the two BLRGs well from radio to $\gamma$-ray
frequencies, and that in both cases the emerging values of the
$\eta$ parameter are similar ($\simeq 0.15-0.35$), being in addition
consistent with the ones following from the analysis by
\citet{gra07}. Importantly, the model when applied to the other
BLRGs analyzed here returns similar values for 
the $\eta$ parameter at the level
of few tens of percent at most (as determined by 
the nuclear radio fluxes), without violating the \F
upper limits (see column 8 in Table\,2 and the SEDs presented in 
Figures\,8). Figure\,4 presents the distribution of the mixing 
parameter $\eta$ emerging from the SED matching. This distribution 
suggests that, even though the particular frequency ranges may be 
dominated by one of the two main emission components, the \emph{total} 
observed luminosities of the nuclear jets in BLRGs constitute on average
not less than $1\%$ of the accretion-related luminosities (median
$\eta \approx 0.10$), and that in the case of a few particular objects the 
jet--to--disk luminosity ratio may even approach unity. An
interesting implication of the above statement, strengthened thanks to
the inclusion of the \F data in our modeling, is that 
one should expect a non-negligible non-thermal emission component in BLRGs
in the MeV energy range, constituting as much as $\sim 1\%-10\%$ of the
total power emitted in the hard X-ray domain. 
If correct, this may be of importance
for understanding the recently debated origin of the extragalactic MeV
background \citep[see the discussions in][]{ino08,aje09}. On the other
hand, the contribution of the jet emission to the total radiative output
of the Seyfert 1 galaxies  are in general very small 
(median $\eta \approx 5 \times 10^{-4}$; see Figure\,4 and the SEDs shown in 
Figure\,9).

To investigate further the collected dataset, 
we plot the optical $[\nu F_{\nu}]_{\rm B}$
versus X-ray $[\nu F_{\nu}]_{\rm 2-10\,keV}$  fluxes for all the sources
included in the sample in Figure\,5 (the
{\it top} panel), with BLRGs denoted by blue filled circles and
Seyferts by red crosses. The {\it bottom} panel of Figure\,4 presents
instead the radio $[\nu F_{\nu}]_{\rm 5\,GHz}$ versus X-ray fluxes. Here,
BLRGs are plotted as blue filled cycles when total radio fluxes are
used, and as cyan open circles when nuclear radio fluxes (obtained predominantly from
VLBI observations where available, supplemented by VLA measurements
in a few cases) are considered. In both panels, large symbols
indicate the two objects detected by \F (3C~111 and 3C~120), while the
medium-size symbols represent Pictor A galaxy which, even though not
formally detected, is characterized by the third-highest TS in the
analyzed two-year \F dataset. These flux-flux plots show that the
optical and X-ray fluxes of BLRGs are roughly linearly correlated, as
expected, though with a substantial (order-of magnitude) scatter. 
In addition, however, it is revealed that 3C~111, 3C~120, and also 
Pictor A, are at the same time the brightest in radio,
but only when the nuclear (and not the total) radio fluxes are taken
into account. Moreover, the nuclear radio fluxes seem well correlated
with the X-ray fluxes for the whole BLRG population. Hence one may
conclude that \F detects preferentially those BLRGs which are
characterized by the brightest radio and X-ray nuclei. This supports the
idea stated previously that the GeV emission of BLRGs is dominated by
the innermost parts of their jets, and is therefore `blazar-like,'
being dependent on the jet luminosity and viewing angle. 
As argued above, the quasar 3C~273, for which the GeV 
luminosity is about 100 times higher than the
nuclear radio luminosity, should provide a reasonably accurate template
for such an emission. Indeed, as illustrated in Figure\,6, the
$\gamma$-ray--to--radio energy flux ratios for the two BLRGs detected by
\F are of the order of $\simeq 100$, while the corresponding upper
limits for all the other objects from the sample (including Seyferts)
are above or much above this value. 
This indicates that the detections of BLRGs
in $\gamma$ rays are at present limited by the sensitivity of the
\F instrument. In fact, from visual inspection of Figure 8 and 9, 
the predicted $\gamma$-ray flux from the template is very close 
to the \F upper limits for Pictor~A, 3C~390.3, 3C~445, 
B3~0309+411B and PKS 2153-69.

Also it is important to note that 
the pc-scale components of the radio jets 
in both 3C~111 and 3C~120 are characterized by apparent superluminal 
motions with maximum velocities 
$\beta_{\rm app} \simeq 5.9$ and $5.3$, respectively, from 
MOJAVE monitoring data \citep{lis09}. The corresponding 
upper limits to the jet viewing angles are thus $\theta_j \lesssim 20^{\circ}$.
Interestingly, mildly relativistic velocities are inferred for 3C~390.3 
and Pictor~A \citep[$\beta_{\rm app}$ up to $\simeq 2.2$ and $1.6$, 
respectively;][]{kel04,tin00}, both of which are good candidates for future \F detections
(see Table\,2 and Figure\,8). These four BLRGs, not coincidentally, 
have the brightest radio/VLBI cores in the analyzed sample, consistent also 
with a relatively large degree of relativistic beaming (see the comment in Appendix\,A). 
Note also that 3C~303 and 3C~382 are other BLRGs with possible
mildly relativistic pc-scale VLBI jets \citep{gio01}.

But what else --- besides the nuclear radio jets --- could be the
source of high-energy emission in the considered objects? Clearly,
large-scale structures (i.e., extended lobes, large-scale jets and
terminal hotspots), which are particularly prominent in BLRGs, are
expected to contribute \emph{at least at some level} to the $\gamma$-ray
emission in the GeV band, since the non-thermal synchrotron continua of
these structures extend to the highest radio frequencies and even 
into the X-ray energies in some cases \citep[see in this context \F detection of the
extended lobes in the low-power radio galaxy
Centaurus~A;][]{cena-lobe}. Such structures are on the other hand almost
absent in Seyfert galaxies, due to much less prominent jet activity in
those objects. Other possible $\gamma$-ray emission sites in both
BLRGs and Seyferts are their accretion disks and disk coronae. Several
related (and rather preliminary) studies presented in the literature, even
though not directly applicable to the types of astrophysical objects
discussed here, indicate that other additional 
processes than those typically considered for generating high-energy photons within the
accreting matter (e.g., proton-proton interactions), may be
relevant in this context. In these scenarios, the resulting disk-related
$\gamma$-ray emission may strongly depend on the main parameters of a
black hole/accretion disk system \citep[such as the black hole mass,
spin, accretion rate, and disk inclination; see,
e.g.,][]{mah97,oka03,nie09}. Hence it is interesting to look into this
issue in more detail for all the sources included in our sample.

Figure\,7 presents therefore the dependence of the $\gamma$-ray
luminosities (or upper limits for such), expressed in the Eddington
units ($L_{\gamma} / L_{\rm Edd}$) or as a ratio of the GeV and X-ray
monochromatic luminosities ($L_{\gamma} / L_{\rm X}$), on  (i) the black 
hole mass $\mathcal{M}_{\rm BH}$ ({\it top left} panel), 
(ii) the accretion ratio $L_{\rm acc} / L_{\rm Edd}$ determined from
the model fitting ({\it top right} panel), 
(iii) the proxy of the radio loudness parameter $L_{\rm R} / L_{\rm B}$ 
({\it bottom left} panel), and (iv) the observed proxy of the accretion 
rate $L_{\rm B} / L_{\rm Edd}$ ({\it bottom right} panel). 
In the figure, BLRGs and Seyferts are denoted by blue and red symbols,
respectively. Large filled circles represent the two BLRGs detected by
\F (3C~111 and 3C~120), while the medium-size open circles represent
Pictor A. In the {\it bottom left} panel, when nuclear radio 
fluxes are used instead of the total radio fluxes, 
BLRGs are denoted by cyan symbols, and Seyferts by magenta symbols. 
A ratio of the GeV and 
X-ray monochromatic luminosities ($L_{\gamma} / L_{\rm X}$) 
expected from a hybrid model is given as dashed lines in the 
{\it bottom right}  panel, for different values of a mixing 
parameter $\eta$ = 10$^{-2}$, 10$^{-1}$, and 1.

Note that the monochromatic luminosity ratio $L_{\rm R} / L_{\rm B}$ is 
simply proportional to the `standard' radio-loudness parameter 
$\mathcal{R}$, and in particular that $L_{\rm R} / L_{\rm B} 
= (\nu_{\rm R} / \nu_{\rm B}) \times \mathcal{R} \sim 10^{-5} \, 
\mathcal{R}$. Also, for the high accretion-rate objects analyzed 
in this paper one expects the accretion luminosity 
$L_{\rm acc} \simeq L_{\rm tot} \simeq 10 \times L_{\rm B}$. 
As shown, the two galaxies which are the brightest in $\gamma$ rays, and
also Pictor~A, are not characterized by any outstanding values of the 
radio loudness, accretion rate, or black hole mass, but are in fact 
quite representative for the other BLRGs included in the sample. 
This supports once again our conclusion that the detected GeV 
emission of BLRGs is related predominantly to nuclear jets, 
and not the other possible nuclear emission components. 
What should be noted for constraining theoretical
models regarding the high-energy emission of accretion disks in AGN is the 
fact that, at least for a number of sources, the \F upper limits
evaluated here already probe the GeV emission of BLRGs and Seyferts down
to the levels of $1\%$ of the X-ray (disk-related) luminosity, or equivalently
$0.01\%$ Eddington luminosity (see Figure\,7).

\acknowledgments 

The \F Collaboration acknowledges generous ongoing support from a number
of agencies and institutes that have supported both the development and
the operation of the LAT as well as scientific data analysis. These
include the National Aeronautics and Space Administration and the
Department of Energy in the United States, the Commissariat \`a
l'Energie Atomique and the Centre National de la Recherche Scientifique
/ Institut National de Physique Nucl\'eaire et de Physique des
Particules in France, the Agenzia Spaziale Italiana and the Istituto
Nazionale di Fisica Nucleare in Italy, the Ministry of Education,
Culture, Sports, Science and Technology (MEXT), High Energy Accelerator
Research Organization (KEK) and Japan Aerospace Exploration Agency
(JAXA) in Japan, and the K.~A.~Wallenberg Foundation, the Swedish
Research Council and the Swedish National Space Board in Sweden.

Additional support for science analysis during the operations phase is
gratefully acknowledged from the Istituto Nazionale di Astrofisica in
Italy and the Centre National d'\'Etudes Spatiales in France.

This research has made use of data from the University of Michigan Radio
Astronomy Observatory which has been supported by the University of
Michigan and by a series of grants from the National Science Foundation,
most recently AST-0607523.

\L.~S. acknowledges the support from the Polish MNiSW through the grant
N-N203-380336. We thank the anonymous referee for critical comments 
which helped to improve the paper.

\appendix

Here we present all the SEDs of the BLRGs (Figure\,8) and Seyfert 1 galaxies 
(Figure\,9) analyzed in this paper which are not detected by \F, together 
with the hybrid model fits described in the main text. The figures illustrate the 
expected level of the GeV emission of each source considered, in comparison
with the current upper limits provided by two-years of \F\
data. Note in particular that the expected 
$\gamma$-ray fluxes of Pictor~A,  3C~390.3 3C~445, B3~0309+411B and 
PKS 2153-69 are close to the \F\ limits derived in this paper (cf., Table\,2). In contrast to the BLRGs (or at least 
the brightest and most beamed examples of such considered in this work), we speculate that the \emph{jet-related} 
non-thermal emission of all the analyzed Seyfert 1 galaxies within the GeV photon 
energy range are in general beyond the level of detectability with \F.

\begin{turnpage}
\begin{deluxetable}{lllrcccccccl}
\tabletypesize{\scriptsize}
\tablecaption{Multiwavelength properties of the analyzed BLRGs and Seyferts}
\tablewidth{0pt}
\tablehead{
\colhead{IAU J2000} & \colhead{name} & \colhead{$z$} & \colhead{$d_{\rm
 L}$} & \colhead{$[\nu F_{\nu}]_{\rm 5\,GHz}^{\rm tot}$} &
 \colhead{$[\nu F_{\nu}]_{\rm 5\,GHz}^{\rm nuc}$} & \colhead{$[\nu
 F_{\nu}]_{\rm B}^{\rm nuc}$} & \colhead{$\Gamma_{\rm X}$} &
 \colhead{$[\nu F_{\nu}]_{\rm 2-10\,keV}$} & \colhead{$[\nu
 F_{\nu}]_{\rm 14-195\,keV}$} &  log $M_{\rm BH}$ &
 \colhead{References$^a$} \\
 \colhead{} & \colhead{} & \colhead{} &
 \colhead{} & \colhead{$\times 10^{-14}$} & \colhead{$\times 10^{-14}$}
 & \colhead{$\times 10^{-12}$} & \colhead{} & \colhead{$\times
 10^{-12}$} & \colhead{$\times 10^{-12}$} & \colhead{} & \colhead{} \\
 \colhead{} & \colhead{} & \colhead{} & \colhead{[Mpc]} &
 \colhead{[erg/cm$^{2}$/s]} & \colhead{[erg/cm$^{2}$/s]} &
 \colhead{[erg/cm$^{2}$/s]} & \colhead{} & \colhead{[erg/cm$^{2}$/s]} &
 \colhead{[erg/cm$^{2}$/s]} & \colhead{[$M_{\odot}$]} & \colhead{}
}
\startdata
 BLRGs & & &  & & & & & & & \\
 & & &  & &  & & & & & \\
\tableline
0040+1003 & 3C~18 & 0.188 & 833.4 & 9.2 & 0.31 & 1.4 & 1.9 & 2.6 &
 -- & 8.92 & 1-5 \\
0313+4120 & B3 0309+411B & 0.134 & 608.1 & 2.5 & 1.6 & 2.2 & 1.9 &
 9.0 & -- & -- & 6-9 \\
0418+3801 & 3C~111 & 0.0485 & 207.1 & 39 & 8.5 & 193 & 1.7 & 37 &
 141 & 8.80 &1,6-7,10-16,25 \\
0433+0521 & 3C~120 & 0.033 & 139.3 & 26 & 19 & 82 & 1.9 & 62 & 119
 & 7.75 &1,10-13,17-24 \\  
0519-4546 & Pictor A & 0.0351 & 148.4 & 77 & 3.8 & 6.8 & 1.7 & 16
 & 38 & 8.70  &1,11$-$13,16,22$-$25\\  
0906+1646 & 3C~215 & 0.412 & 2183 & 2.1 & 0.082 & 1.5 & 1.7 & 2.3 & --
 & 8.80 & 11,29$-$31 \\
0947+0725 & 3C~227 & 0.0858 & 376.6 & 13 & 0.11 & 6.2 & 1.5f &
 1.0$^b$ & 27 & 8.90  & 13,32$-$35 \\
1443+5201 & 3C~303 & 0.141 & 642.9 & 4.7 & 0.75 & 0.7 & 1.9 & 2.0
 & -- & 8.00 &  11,25,35,36 \\
1722+2436 & RGB J1722+246 & 0.175 & 815.6 & 0.17 & 0.14 &
 7.1 & 1.8 & 1.0 & -- & -- &  1,3,37 \\   
1835+3241 & 3C~382 & 0.0579 & 249.0 & 11 & 0.94 & 17 & 1.7 & 46 &
 84 & 8.90  &   1,9$-$13,22,25,38,39\\   
1842+7946 & 3C~390.3 & 0.0561 & 240.9 & 22 & 2.0 & 31 & 1.7 & 34 & 110 &
 8.80& 1,6,7,9$-$11,13,17, \\   
 &  &  &  &  &  &  &  &  &  &
 & 19,20,22,25,40,41 \\   
2022+1001 & 3C~411 & 0.467 & 2536 & 4.5 & 0.39 & 0.6 & 1.8 & 0.8 & -- & --
 & 2,11,31,75 \\ 
2042+7508 & 4C~74.26 & 0.104 & 462.4 & 1.7 & 0.5 & 74 & 1.8 & 20 & 50 & 9.37
 & 6,7,9,13,22,42,43\\ 
2114+8204 & S5~2116+81 & 0.084 & 368.2 & 1.2 & 0.49 & 17 & 1.9 &
 15 & 42 & 8.12 & 1,6,7,9,13,43,44\\
2124+5058 & 4C~50.55 & 0.020 & 83.6 & 5.0 & 1.0$^d$ & --$^c$ & 1.4
 & 51 & 178 & -- &  6,7,9,13,45\\ 
2157-6941 & PKS 2153-69 & 0.0283 & 119 & 60 & 1.5 & 44 & 1.8 & 7.2 & --
 & -- &1,46,47\\
 2223-0206 & 3C~445 & 0.0562 & 241.4 & 10 & 0.41 & 2.9 & 1.6 & 15 &
 44 & 8.00 &1,9$-$11,13,22, \\ 
  &  &  &  &  &  &  &  &  &
  &  & 25,48$-$51 \\ 
2254+1136 & PKS 2251+11 & 0.326 & 1656 & 3.0 & 0.01 & 11 & 1.1 &
 1.5 & -- & 8.93 &  11,31,52,53 \\ 
\tableline
 & & &  & &  & & & & & \\
Seyferts & & &  & & & & & & & \\
 & & &  & &  & & & & & \\
\tableline
0006+2012 & Mrk~335 & 0.0258 & 114.2 & 0.016 & 0.016 & 5.1 &
 2.0 & 13.5 & 24.7 & 6.80 & 13,22,25,54$-$60,76\\
0123-5848 & Fairall~9 & 0.047 & 214.1 & $<$0.025 & $<$0.025 & 145 &
 1.7 & 20.5 & 50.7 & 7.90 & 13,22,25,54,61$-$63,76,77\\
0516-0008 & Ark~120 & 0.0327 & 144.2 & 0.062 & 0.015 & 201 &
 2.2 & 32.5 & 70.8 & 8.30 & 13,22,25,54,64,65,76\\
0925+5217 & Mrk~110 & 0.0353 & 158.3 & 0.027 & 0.011 & 8.4 &
 1.7 & 28.0 & 61.5 & 6.70 & 13,22,25,54,64$-$66,76\\
1139-3744 & NGC~3783 & 0.00973 & 38.5 & 0.12 & 0.065 & 89 &
 1.6 & 49.0 & 195 & 7.00 & 9,13,22,25,54, \\
 & & & & & &  &
 & & & & 61,64,67,76 \\
1239-0520 & NGC~4593 & 0.009 & 39.5 & 0.013 & 0.0081 & 27 & 1.8
 & 36.5 & 97.9 & 6.90 & 13,22,25,54,68$-$71,76 \\
1349-3018 & IC4329A & 0.0161 & 70.2 & 0.18 & 0.17 & 34 & 1.7 &
 90.0 & 331 & 6.70 & 9,22,25,54,61,64,72,76 \\
2044-1043 & Mrk~509 & 0.0344 & 154.1 & 0.051 & 0.026 & 140 &
 1.6 & 44.0 & 94.4 & 7.80 & 9,22,25,54,61,68,71,76 \\
2303+0852 & NGC~7469 & 0.0163 & 71.4 & 0.39 & 0.30 & 10 & 1.8
 & 31.5 & 66.6 & 6.80 & 9,25,54,61,64,74,76 \\
\tableline
\enddata 
\tablenotetext{a}{References: 
[1] \citet{gra06}; [2] \citet{hew91}; [3] \citet{lau97};
[4] \citet{mar04}; [5] \citet{bri94}; [6] \citet{mol08};
[7] \citet{mol09}; [8] \citet{hen95}; [9] \citet{bir07}; 
[10] \citet{woz98}; [11] \citet{sam99}; [12] \citet{era00}; 
[13] \citet{tue10}; [14] \citet{tur89}; [15] \citet{lin87}; 
[16] \citet{lew05}; [17] \citet{gli03}; [18] \citet{kat07}; 
[19] \citet{hir00}; [20] \citet{fom00}; [21] \citet{zho10}; 
[22] \citet{fuk11}; [23] \citet{ogl05}; [24] \citet{bal04}; 
[25] \citet{sik07}; [26] \citet{tin03}; [27] \citet{tin00};
[28] \citet{era98}; [29] \citet{hou02}; [30] \citet{lab06};
[31] \citet{ree00}; [32] \citet{cra95}; [33] \citet{har07};
[34] \citet{aje08}; [35] \citet{kat03}; [36] \citet{gio90};
[37] \citet{don05}; [38] \citet{gio94}; [39] \citet{sam11};
[40] \citet{gli09}; [41] \citet{sam09}; [42] \citet{pea92};
[43] \citet{wan09}; [44] \citet{tay96}; [45] \citet{taz10};
[46] \citet{tin02}; [47] \citet{you05}; [48] \citet{pre83}; 
[49] \citet{gran07}; [50] \citet{sam07}; [51] \citet{sam98};
[52] \citet{kel89}; [53] \citet{liu06}; [54] \citet{kas05};
[55] \citet{ede87}; [56] \citet{gal06}; [57] \citet{ino07}; 
[58] \citet{lar08}; [59] \citet{bia01}; [60] \citet{one07}; 
[61] \citet{shi06}; [62] \citet{sch09}; [63] \citet{gon01};
[64] \citet{nan07}; [65] \citet{saz07}; [66] \citet{das06};
[67] \citet{der02}; [66] \citet{bec06}; [69] \citet{rey04};
[70] \citet{bre07}; [71] \citet{mar09}; [72] \citet{ste05};
[73] \citet{ros04}; [74] \citet{blu03}: [75] \citet{nef95};
[76] \citet{ho02}; [77] \citet{ver91}.}
\tablenotetext{b}{Flux estimated from the counts rate of {\it Chandra} ACIS CCD chip using \textsc{PIMMS} and assuming the X-ray photon index $\Gamma_{\rm X} = 1.5$.}
\tablenotetext{c}{Unreliable extinction estimate due to the location of the source at low Galactic latitude $| b | < 5$\,deg.}
\tablenotetext{d}{5 GHz flux was obtained from our new VLBA observations 
obtained on 2009 Feb 7 in program BC~185 (PI: C. C. Cheung).}
\end{deluxetable}
\end{turnpage}

\begin{deluxetable}{lccccccr}
\tabletypesize{\scriptsize}
\tablecaption{Results of the \F data analysis}
\tablewidth{0pt}
\tablehead{
\colhead{name} & \colhead{TS} & \colhead{$\Gamma_{\gamma}$} &
 \colhead{$F_{\rm >0.1\,GeV}$} & \colhead{$[\nu F_{\nu}]_{\rm
 0.1-10\,GeV}$} & \colhead{$\log L_{\gamma}$} & \colhead{$\log L_{\rm acc}$$^a$}  & \colhead{$\eta$$^b$}  \\
\colhead{} & \colhead{} & \colhead{} & \colhead{[$10^{-9}$\,ph\,cm$^{-2}$\,s$^{-1}$]} & \colhead{[$10^{-12}$\,erg\,cm$^{-2}$\,s$^{-1}$]} &
 \colhead{[erg\,s$^{-1}$]} & \colhead{[erg\,s$^{-1}$]}  & \colhead{} }
\startdata
BLRGs & & & & & \\
 & & & & &\\
\tableline
3C~18 & 1.5 & 2.5f$^c$ & $<$ 9.2 & $<$ 4.0 & $<$ 44.6 & 45.3  & 0.070  \\
B3 0309+411B & $<$ 1 & 2.5f & $<$ 8.9 & $<$ 3.9 & $<$ 44.3 & 45.5 & 0.12 \\
3C~111 & 31 & 2.7$\pm$0.2 & 35$\pm$12 &  15 & 43.9 & 45.0 & 0.35 \\
3C~120 & 34 & 3.0$\pm$0.3 & 37$\pm$14 &  16 & 43.6 & 44.9 & 0.15\\
Pictor A & 20 & 2.5f & $<$ 15 & $<$ 6.3 & $<$ 43.2 & 44.5 & 0.25\\
3C~215   & 13 & 2.5f & $<$ 14 & $<$ 6.1 & $<$ 45.6 & 45.9 & 0.034 \\
3C~227 & $<$ 1 & 2.5f & $<$ 5.0 & $<$ 2.2 & $<$ 43.6 & 44.9 & 0.017\\
3C~303 & 2.0 & 2.5f & $<$ 5.9 & $<$ 2.6 & $<$ 44.1 & 45.0 & 0.20\\ 
RGB J1722+246 & $<$ 1 & 2.5f & $<$ 38 & $<$ 16 & $<$ 45.1 & 45.0 & 0.058\\
3C~382 & 2.8  & 2.5f & $<$ 12 & $<$ 5.2 & $<$ 43.6 & 45.1 & 0.025\\   
3C~390.3 & 3.2 & 2.5f & $<$ 7.4 & $<$ 3.2 & $<$ 43.4 & 45.0& 0.075\\
3C~411 & 8.4 &  2.5f & $<$ 18 & $<$ 7.6 &  $<$ 45.8 & 45.7 & 0.17\\
4C~74.26 & 3.1 & 2.5f & $<$ 11 & $<$ 4.7 & $<$ 44.1 & 45.6 & 0.015\\
S5~2116+81 & 1.6 & 2.5f & $<$ 5.9 & $<$ 2.6 & $<$ 43.6 & 45.3 & 0.023\\
4C~50.55 & $<$ 1 & 2.5f & $<$ 180 & $<$ 77 & $<$ 43.8 & 44.4 & 0.024\\
PKS 2153-69 & 1.4$^d$ & 2.5f & $<$ 10 & $<$ 4.4 & $<$ 42.9 & 44.2 & 0.10\\
3C~445 & $<$ 1 & 2.5f & $<$ 2.1 & $<$ 0.90 & $<$ 42.8 & 44.8 & 0.022\\
PKS 2251+11 & $<$ 1 & 2.5f & $<$ 5.4 & $<$ 2.3 & $<$ 44.9 & 45.4 & 0.010\\
\tableline
 & & & & \\
Seyferts  & & & & &\\
 & & & & &\\
\tableline
Mrk~335 & 8.9 & 2.5f & $<$ 11 & $<$ 4.8 & $<$ 42.9 & 44.6 & 
 2.6$\times$10$^{-4}$ \\
Fairall~9 & $<$ 1 & 2.5f & $<$ 1.9 & $<$ 0.81 & $<$ 42.6 & 45.0 & 
$<$ 4.0$\times$10$^{-4}$  \\
Ark~120 & $<$ 1 & 2.5f & $<$ 2.8 & $<$ 1.2 & $<$ 42.5 & 44.9 & 
 2.3$\times$10$^{-4}$  \\
Mrk~110 & 3.9 & 2.5f & $<$ 6.9 & $<$ 3.0 & $<$ 42.9 & 44.8 & 
 3.3$\times$10$^{-4}$  \\
NGC~3783 & $<$ 1 & 2.5f & $<$ 3.2 & $<$ 1.4 & $<$ 41.5 & 44.1 & 
 4.6$\times$10$^{-4}$  \\
NGC~4593 & 2.5 & 2.5f & $<$ 8.9 & $<$ 3.9 & $<$ 41.8 & 43.9 & 
 1.3$\times$10$^{-4}$  \\
IC4329A & 15 & 2.5f & $<$ 19 & $<$ 8.1 & $<$ 42.7  & 44.7 & 
 1.0$\times$10$^{-3}$ \\
Mrk~509 & $<$ 1 & 2.5f & $<$ 2.7 & $<$ 1.2 & $<$ 42.5 & 45.1 & 
 3.2$\times$10$^{-4}$ \\
NGC~7469 & $<$ 1 & 2.5f & $<$ 7.1 & $<$ 3.1 & $<$ 42.3 & 44.8 & 
 1.2$\times$10$^{-3}$  \\
\tableline
\enddata 
\tablenotetext{a}{Accretion luminosity derived from the SED 
fitting, assuming hybrid model discussed in the paper.}
\tablenotetext{b}{Mixing parameter described in the text, 
derived from the SED fitting.}
\tablenotetext{c}{Photon spectral index was fixed at 2.5.}
\tablenotetext{d}{A nearby LAT source was found 0.27 deg apart from
 PKS~2153-69 with TS of 44. We treat this as a background source since
 its separation from the target is more than 2 times larger than its 
 95 $\%$ $\gamma$-ray localization error radius.}
\end{deluxetable}

\begin{figure}
\begin{center}
\includegraphics[angle=0,scale=0.4]{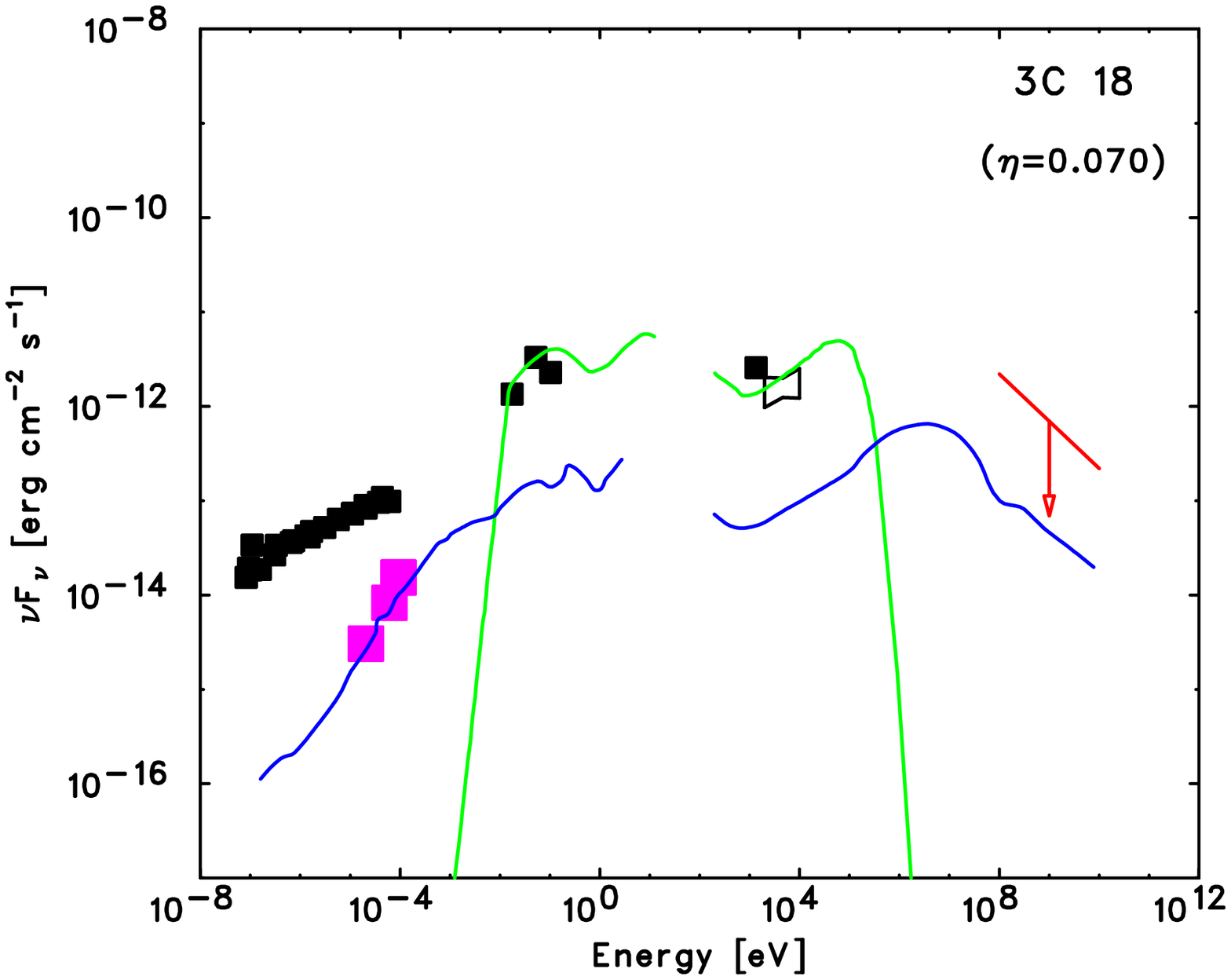}
\includegraphics[angle=0,scale=0.4]{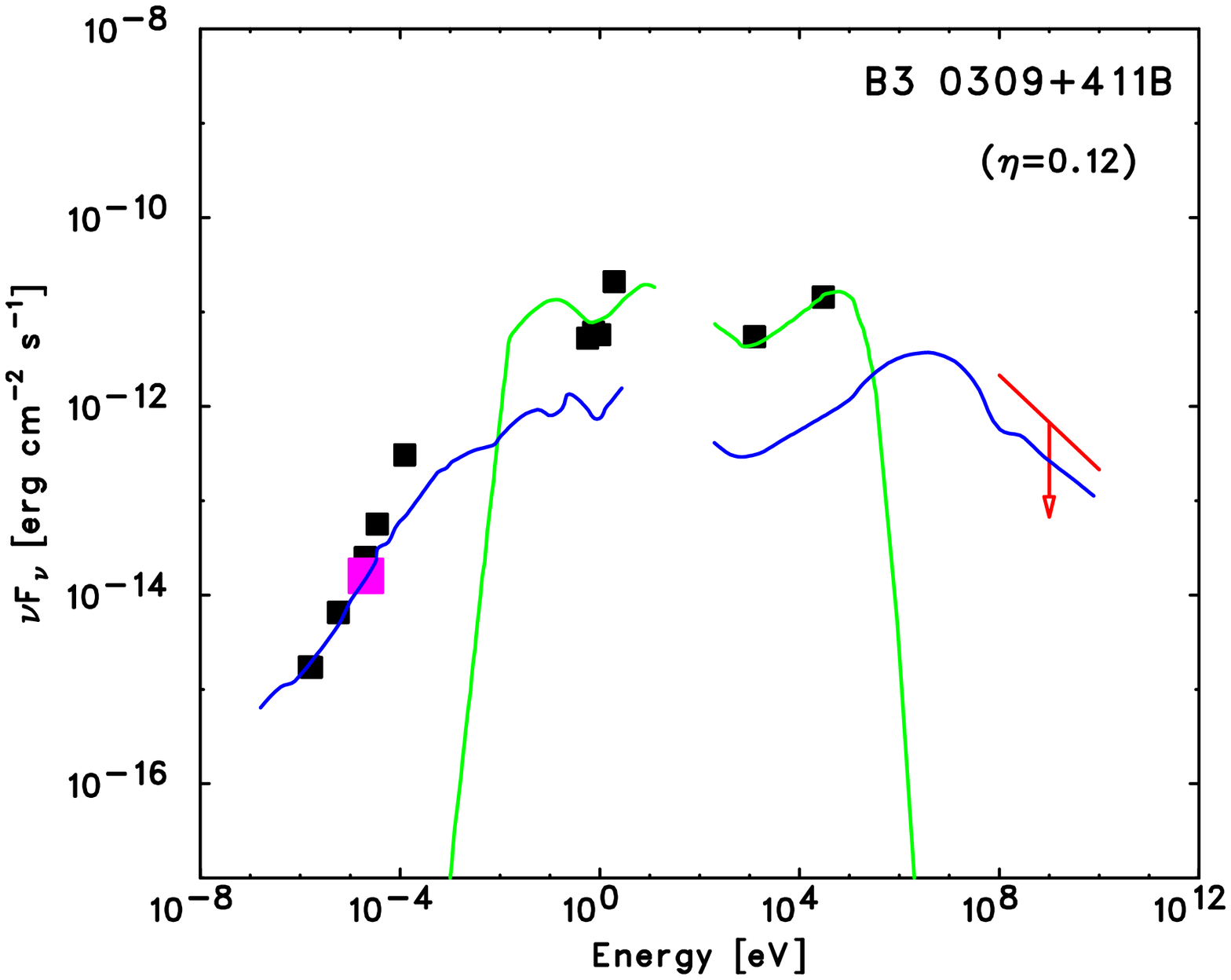}
\includegraphics[angle=0,scale=0.4]{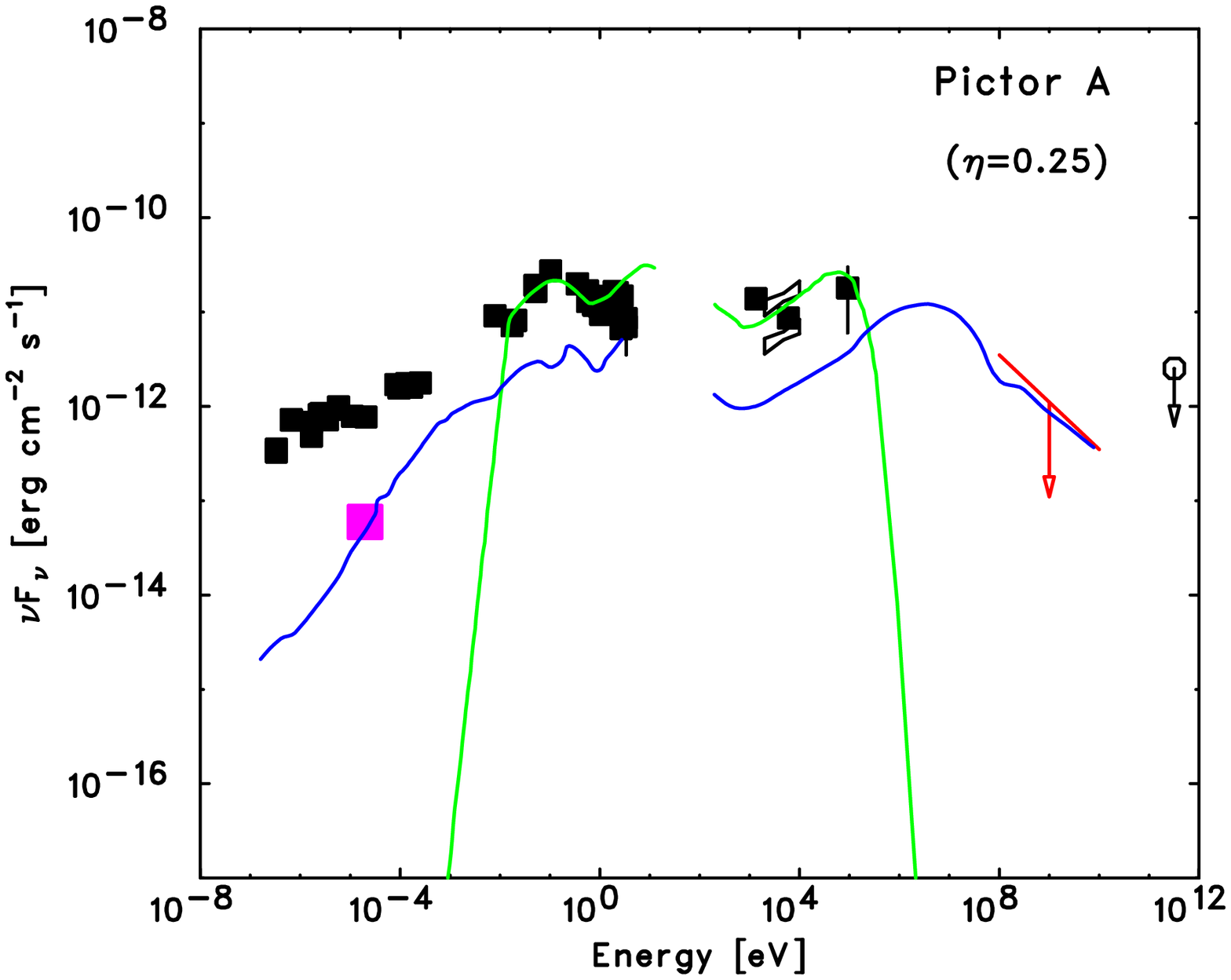}
\includegraphics[angle=0,scale=0.4]{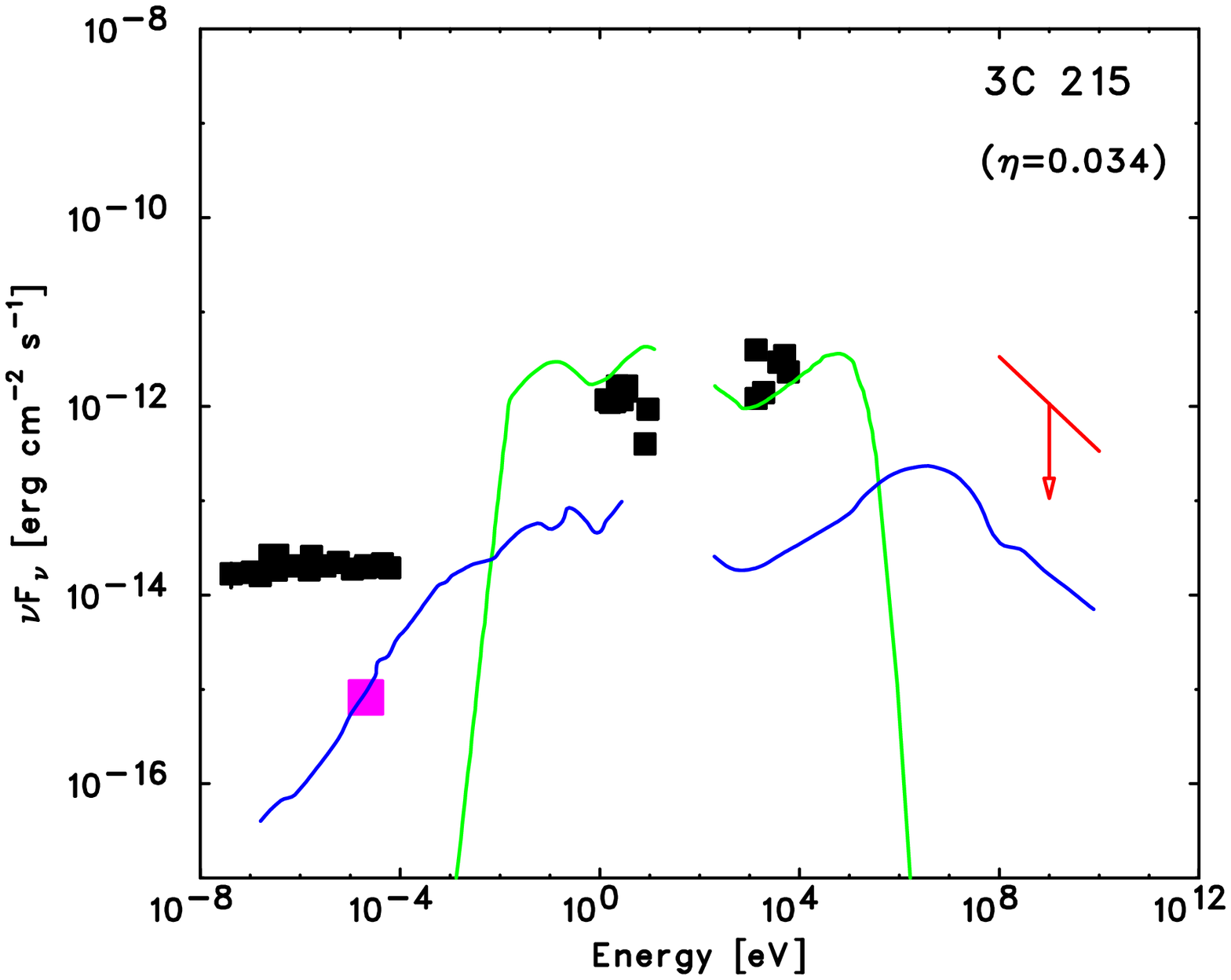}
\includegraphics[angle=0,scale=0.4]{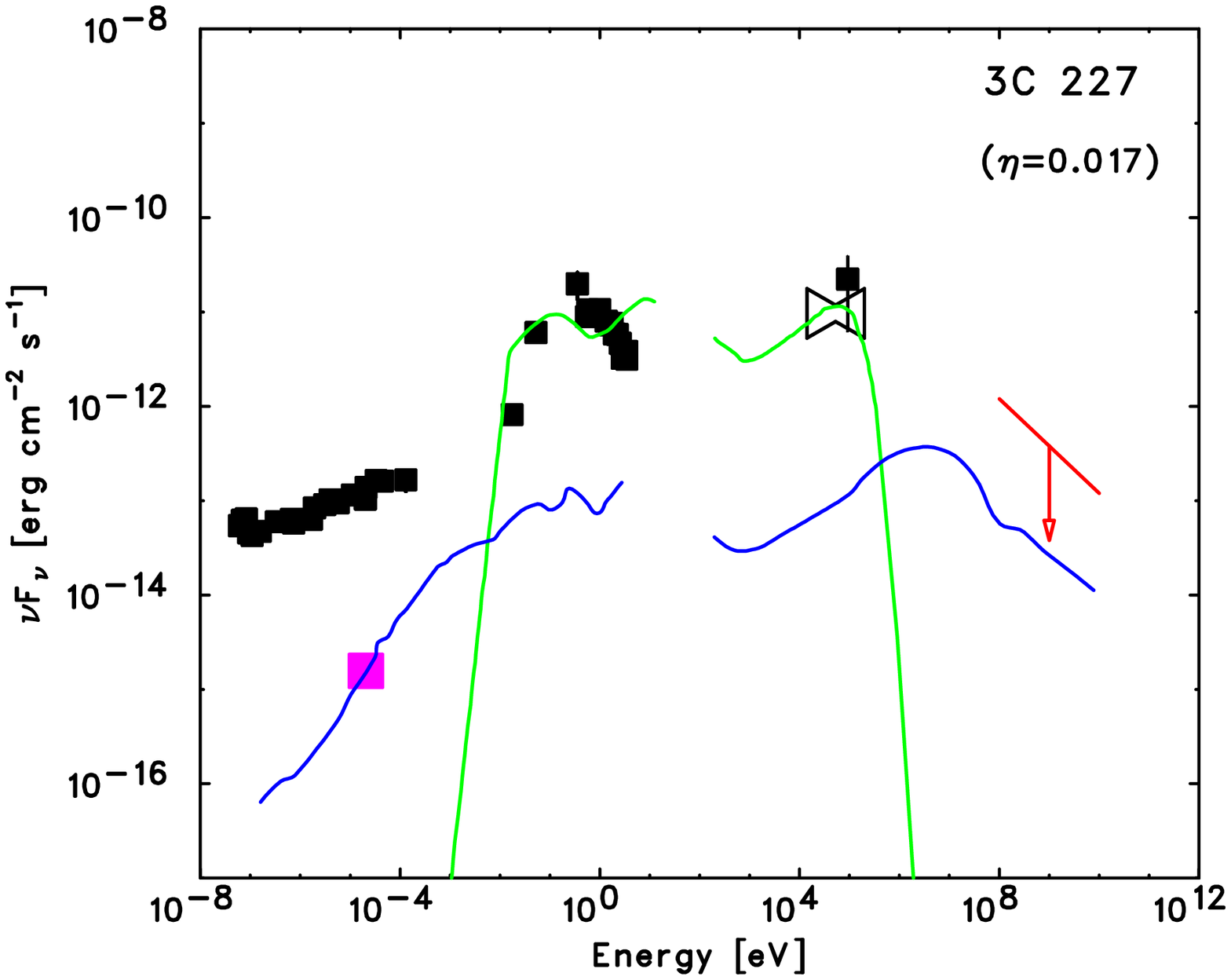}
\includegraphics[angle=0,scale=0.4]{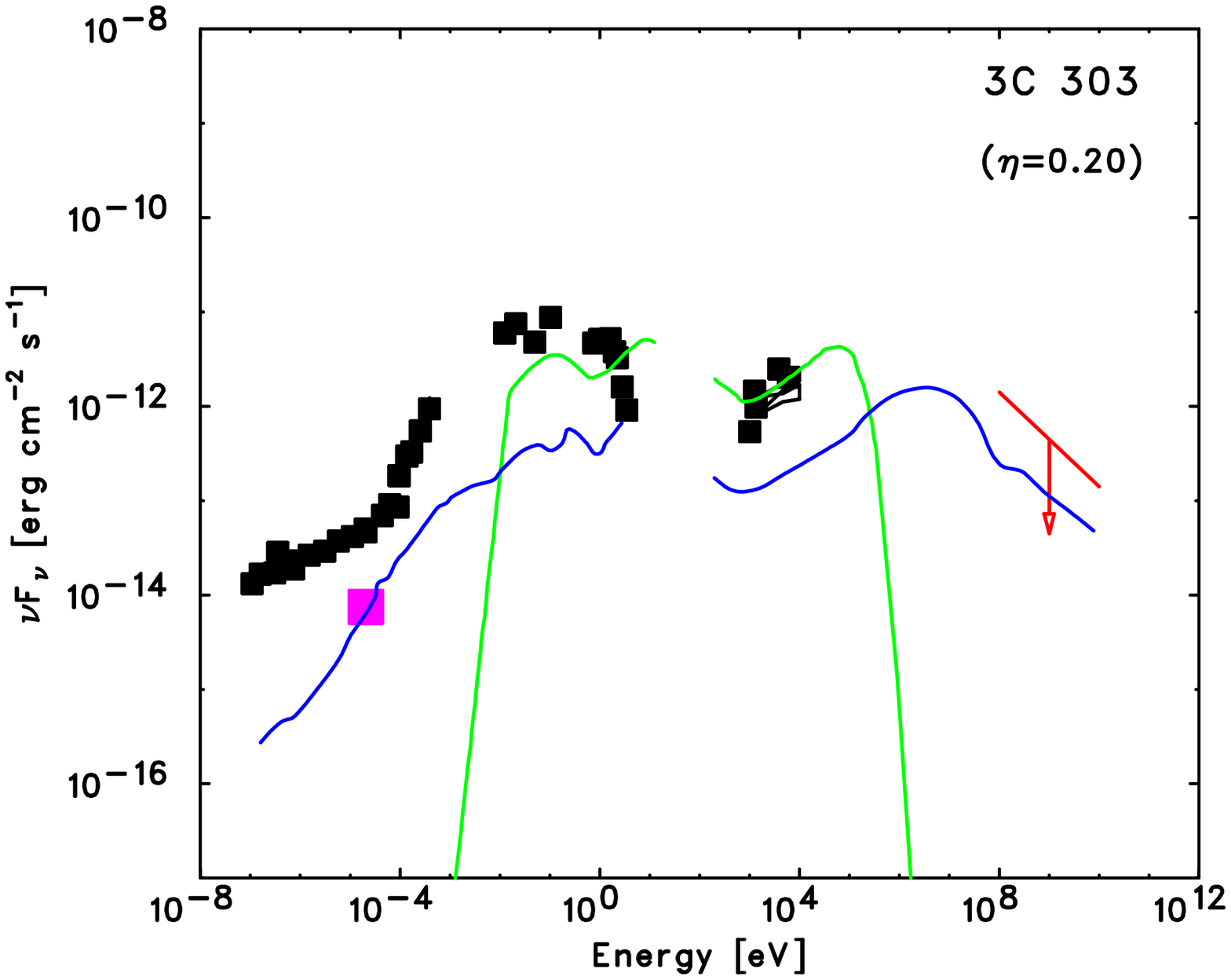}
\caption{Broad-band SEDs of the BLRGs which are $not$ detected at high
 significance in the GeV photon energy range. \F upper limits are indicated by red arrows. Black squares represent the historical data from NED. Magenta squares denote the $5$\,GHz radio fluxes of the unresolved nuclei. 
The green curves correspond to the template of the accretion-related
 Seyfert-type emission \citep[from][]{kor99}, matched to the
 infrared--to--X-ray continuum of each source. The blue curves correspond
 to the broad-band spectrum of the quasar 3C~273 \citep[from][]{sol08},
 used here as a template of the jet-related emission and scaled to match
 the radio fluxes for each source. The mixing parameter $\eta$ for the
 phenomenological hybrid model discussed in \S\,4 is given in each
 panel.}
\end{center}
\end{figure}

\addtocounter{figure}{-1}
\begin{figure}
\begin{center}
\includegraphics[angle=0,scale=0.4]{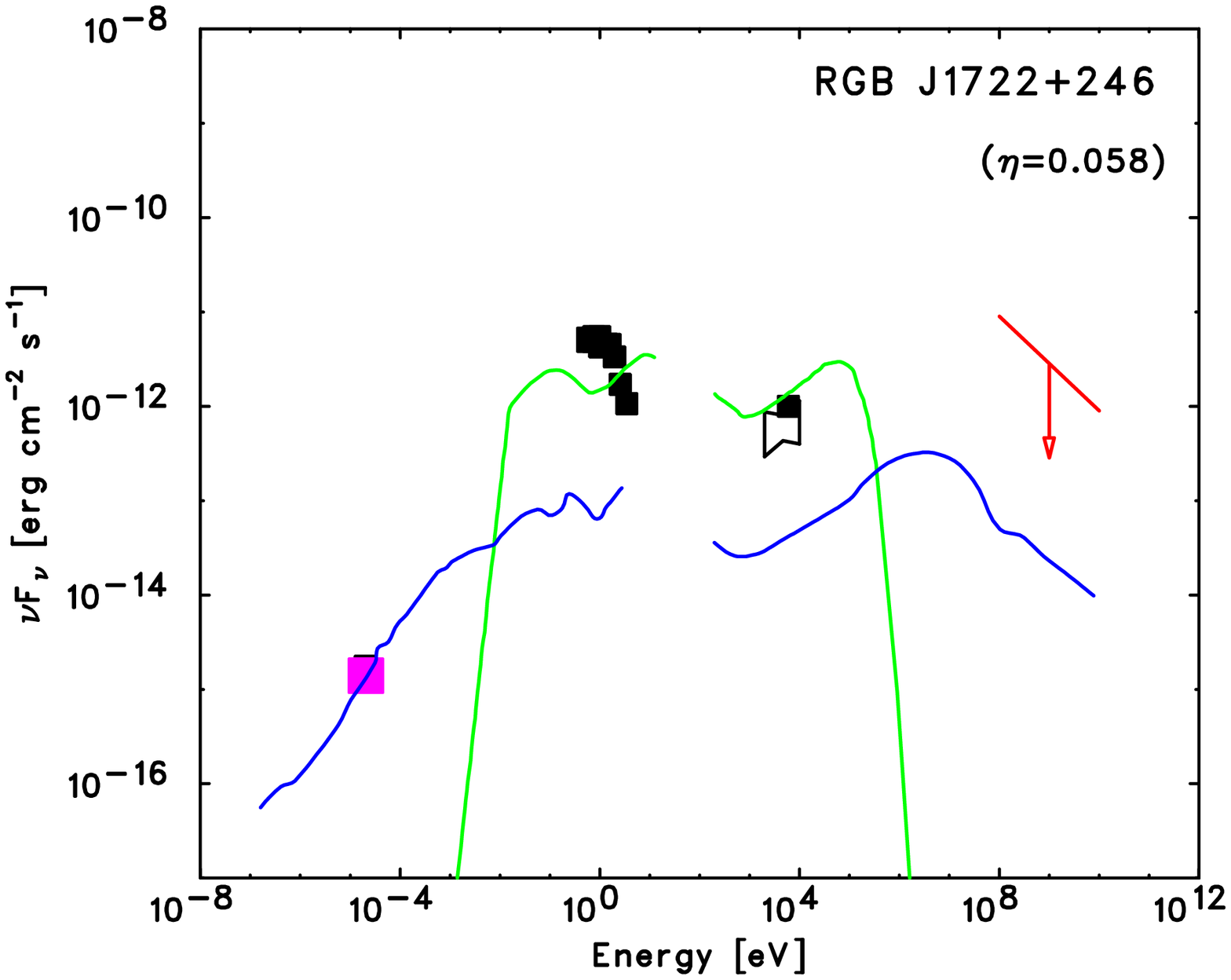}
\includegraphics[angle=0,scale=0.4]{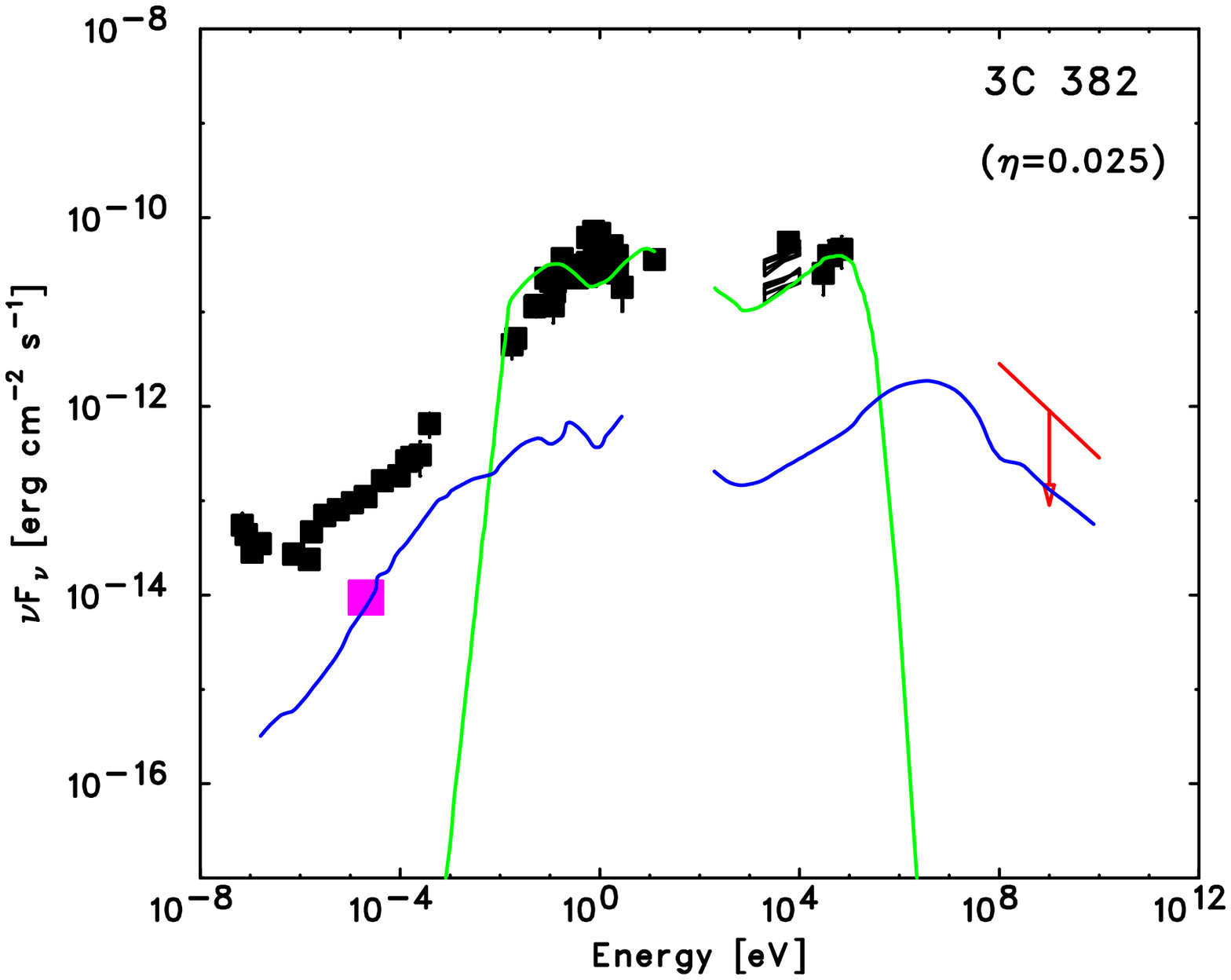}
\includegraphics[angle=0,scale=0.4]{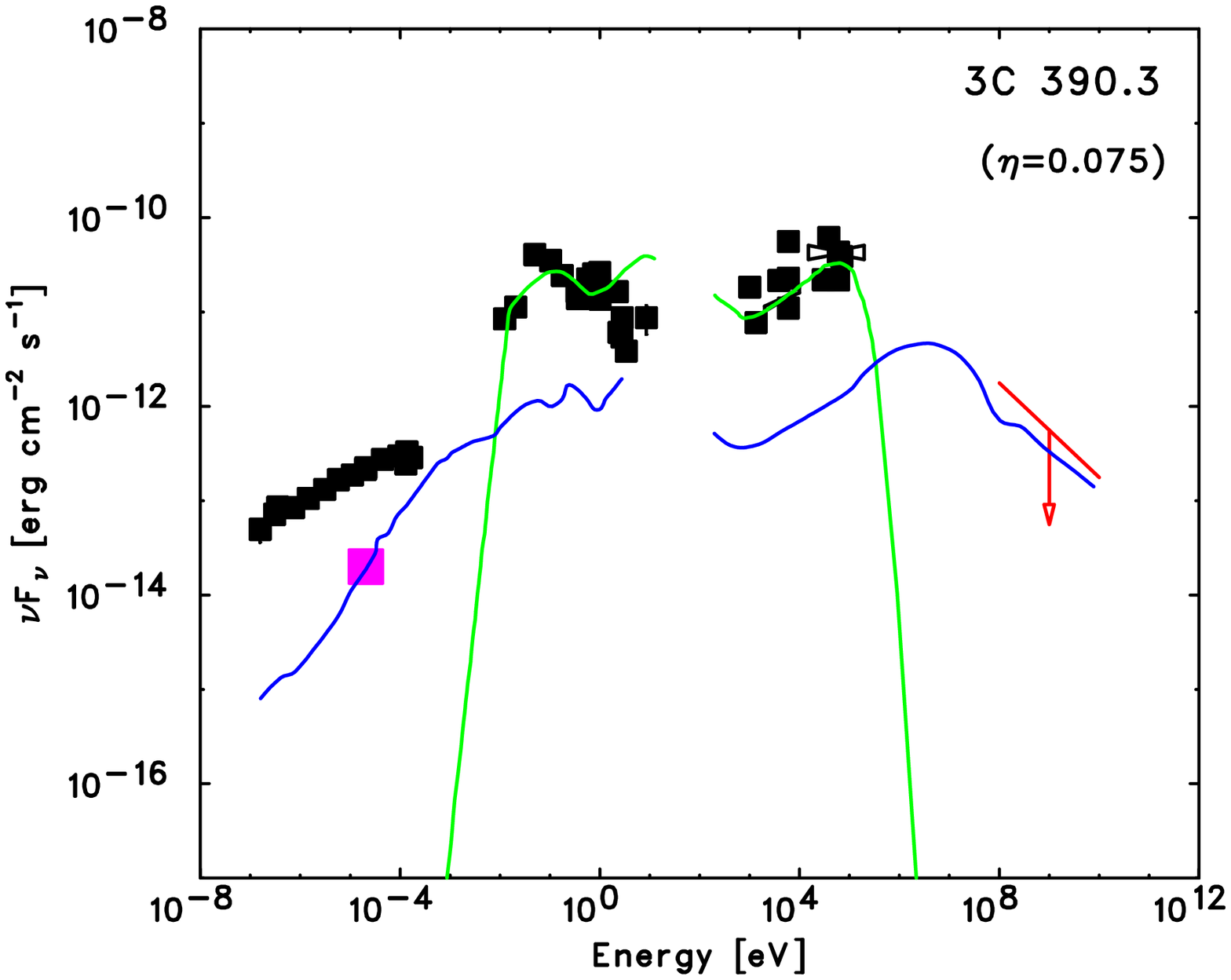}
\includegraphics[angle=0,scale=0.4]{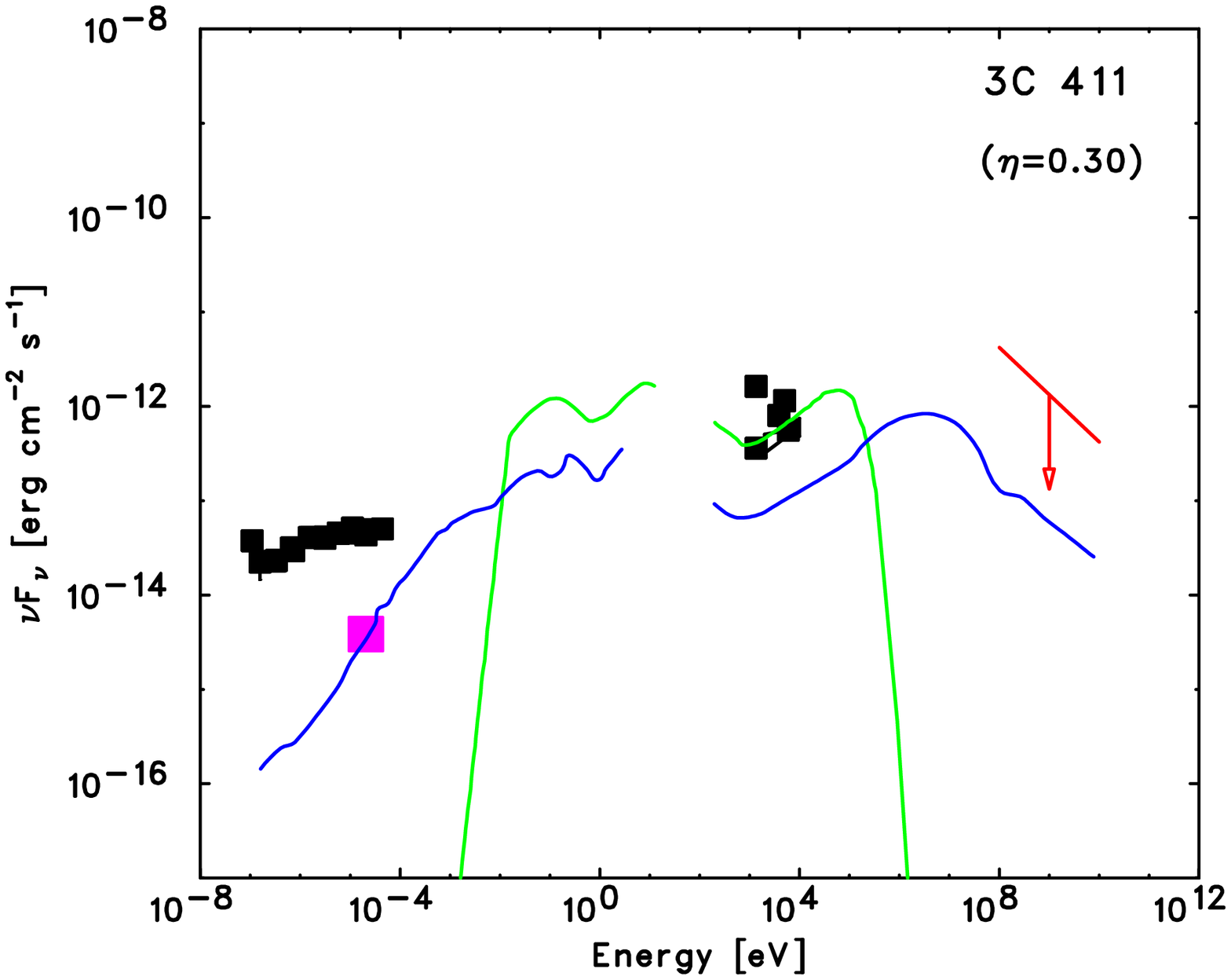}
\includegraphics[angle=0,scale=0.4]{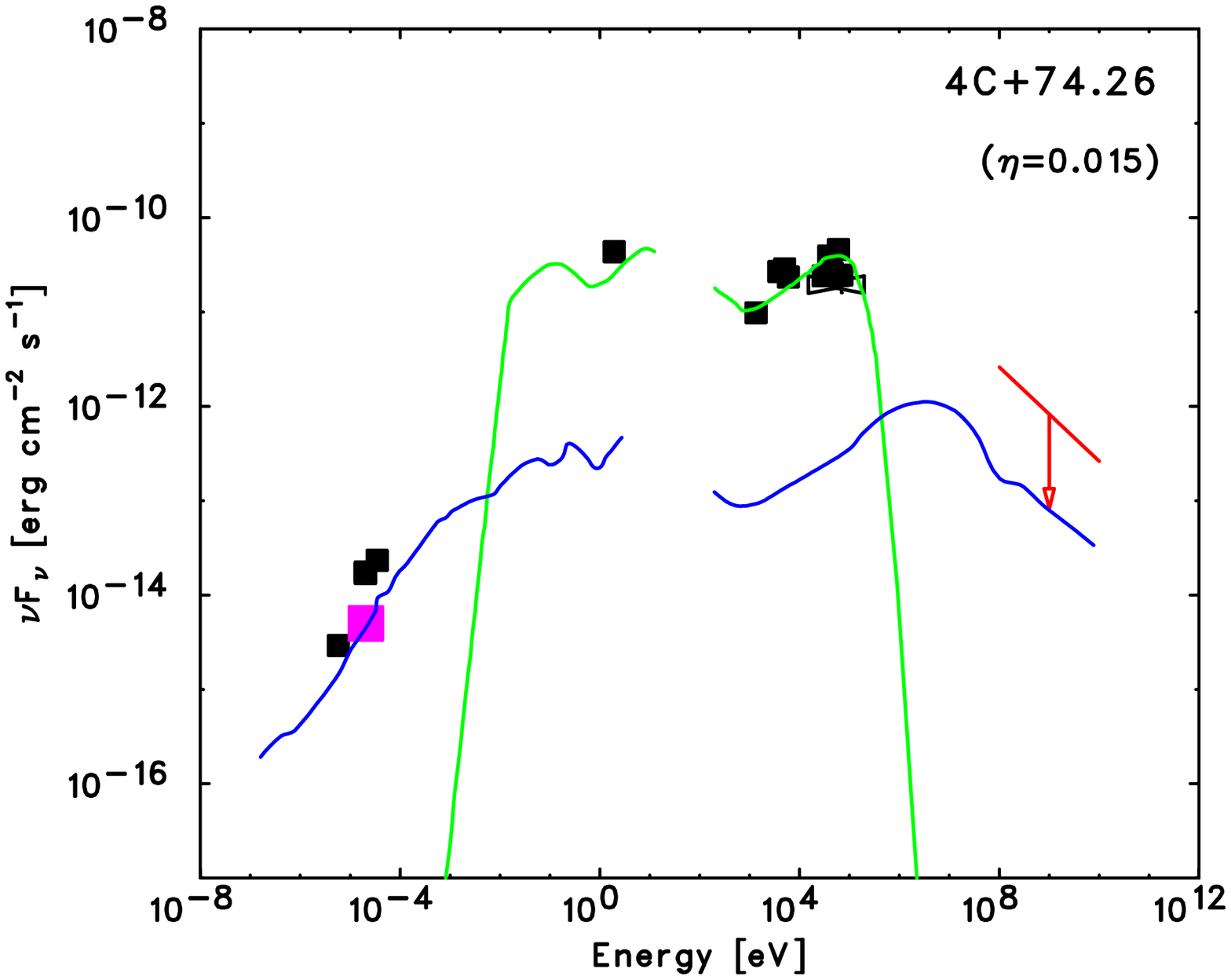}
\includegraphics[angle=0,scale=0.4]{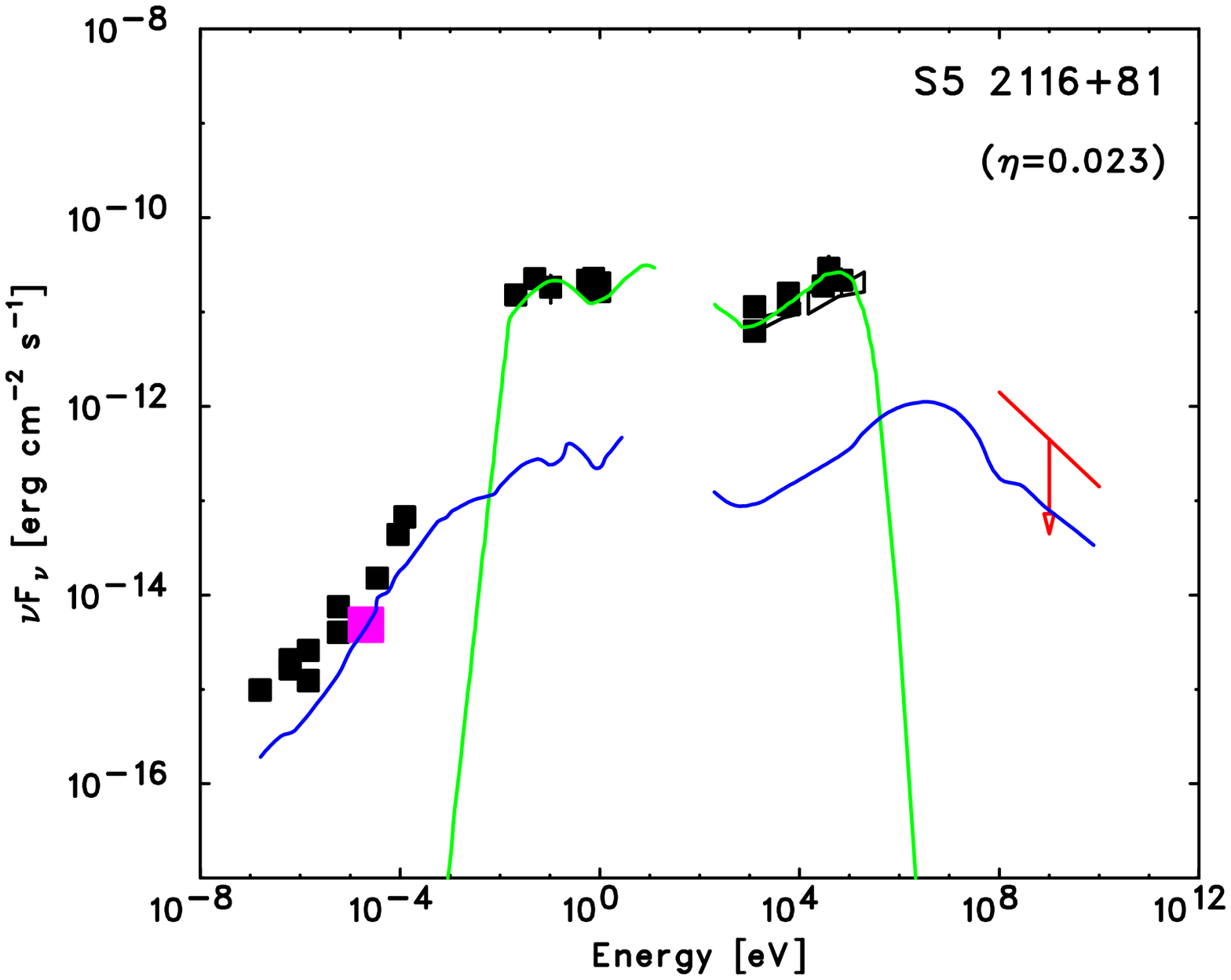}
\caption{$- continued$.}
\end{center}
\end{figure}

\addtocounter{figure}{-1}
\begin{figure}
\begin{center}
\includegraphics[angle=0,scale=0.4]{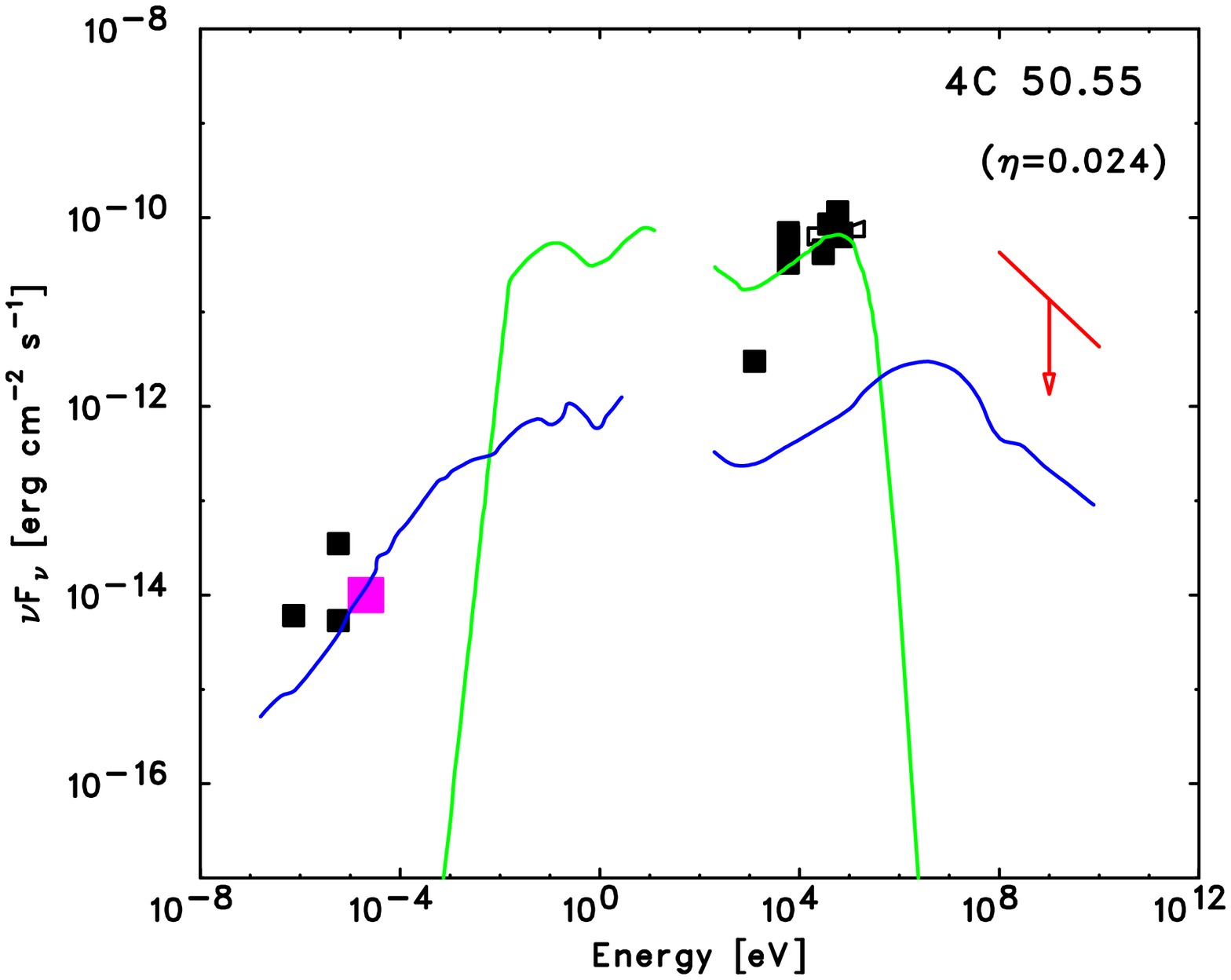}
\includegraphics[angle=0,scale=0.4]{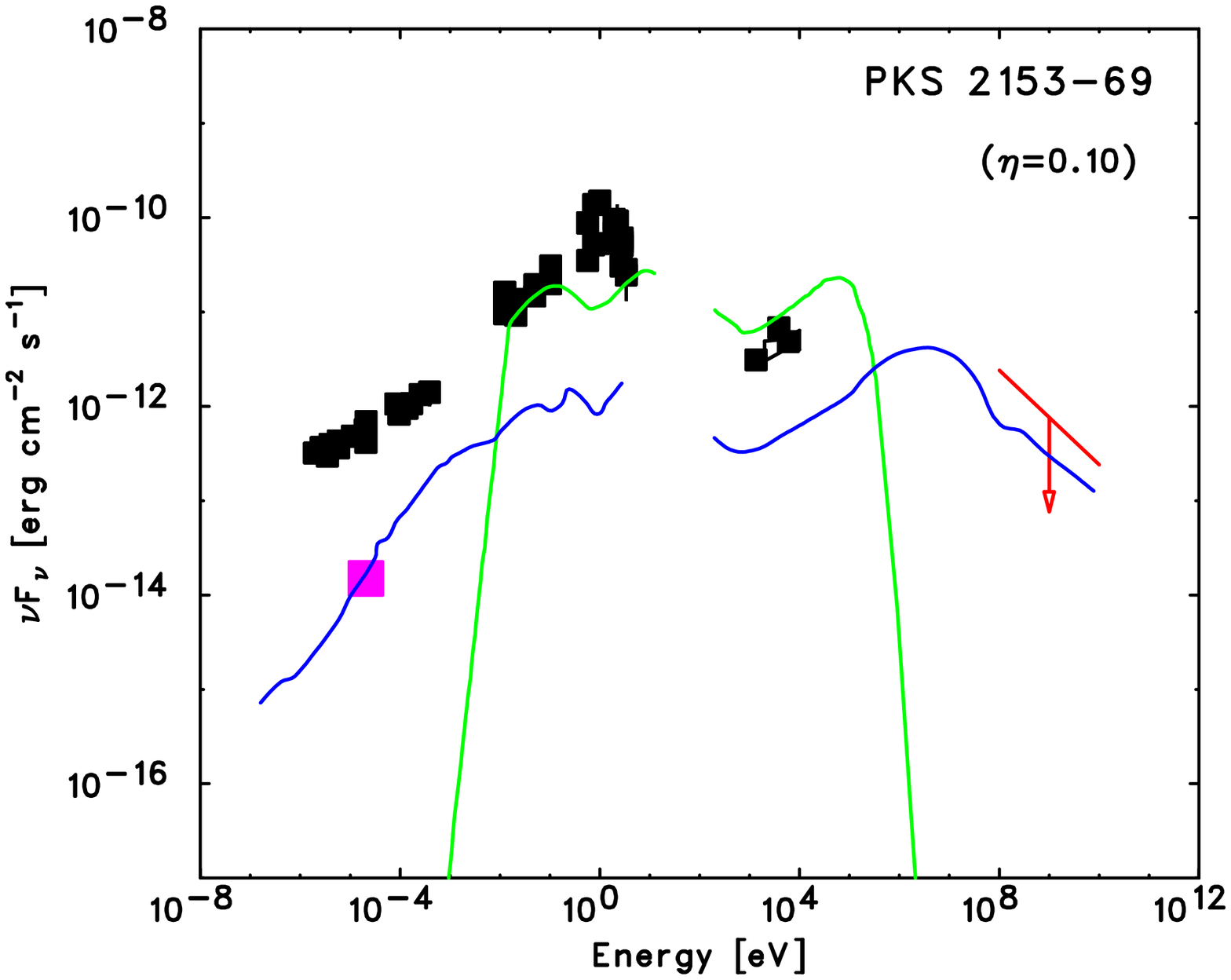}
\includegraphics[angle=0,scale=0.4]{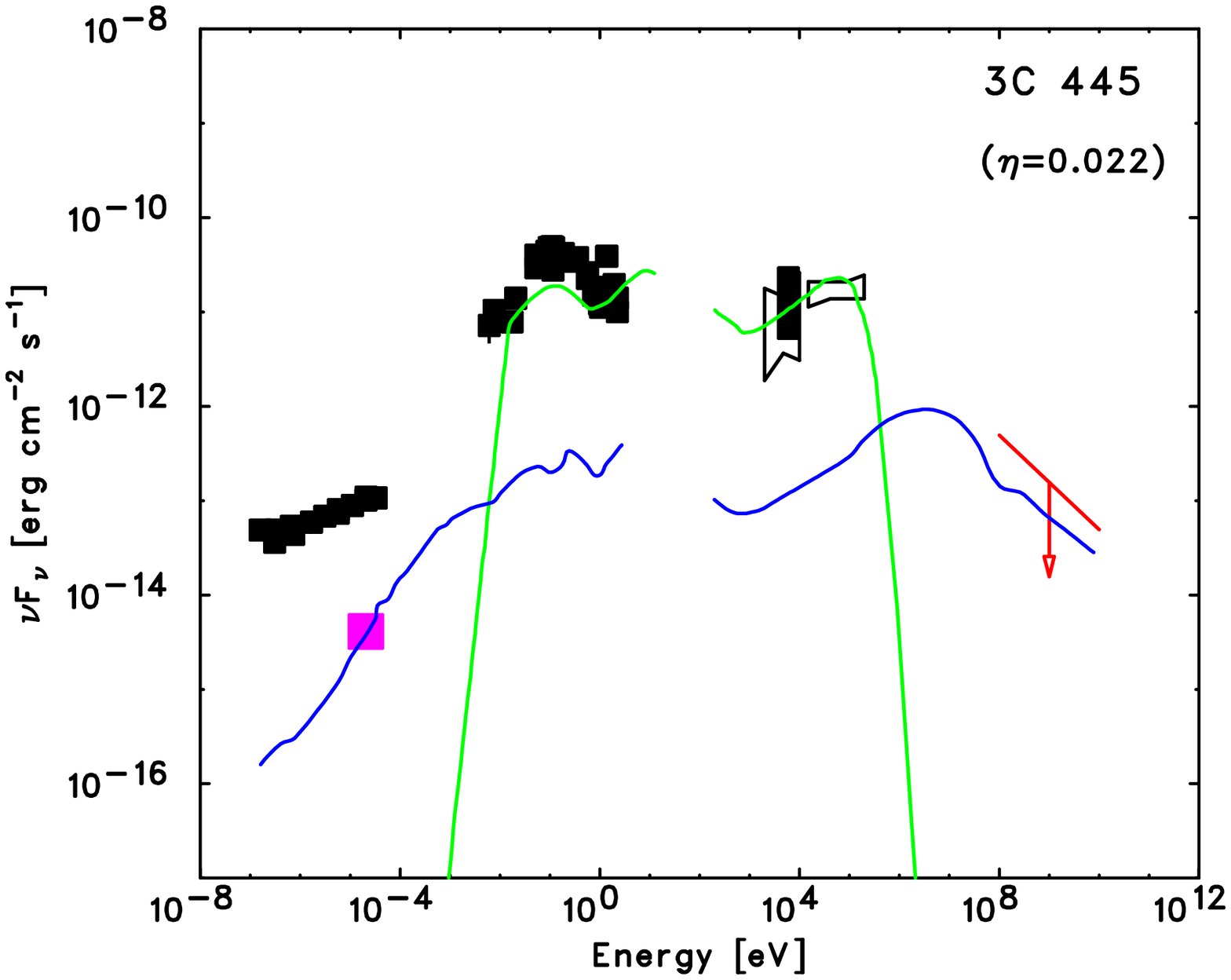}
\includegraphics[angle=0,scale=0.4]{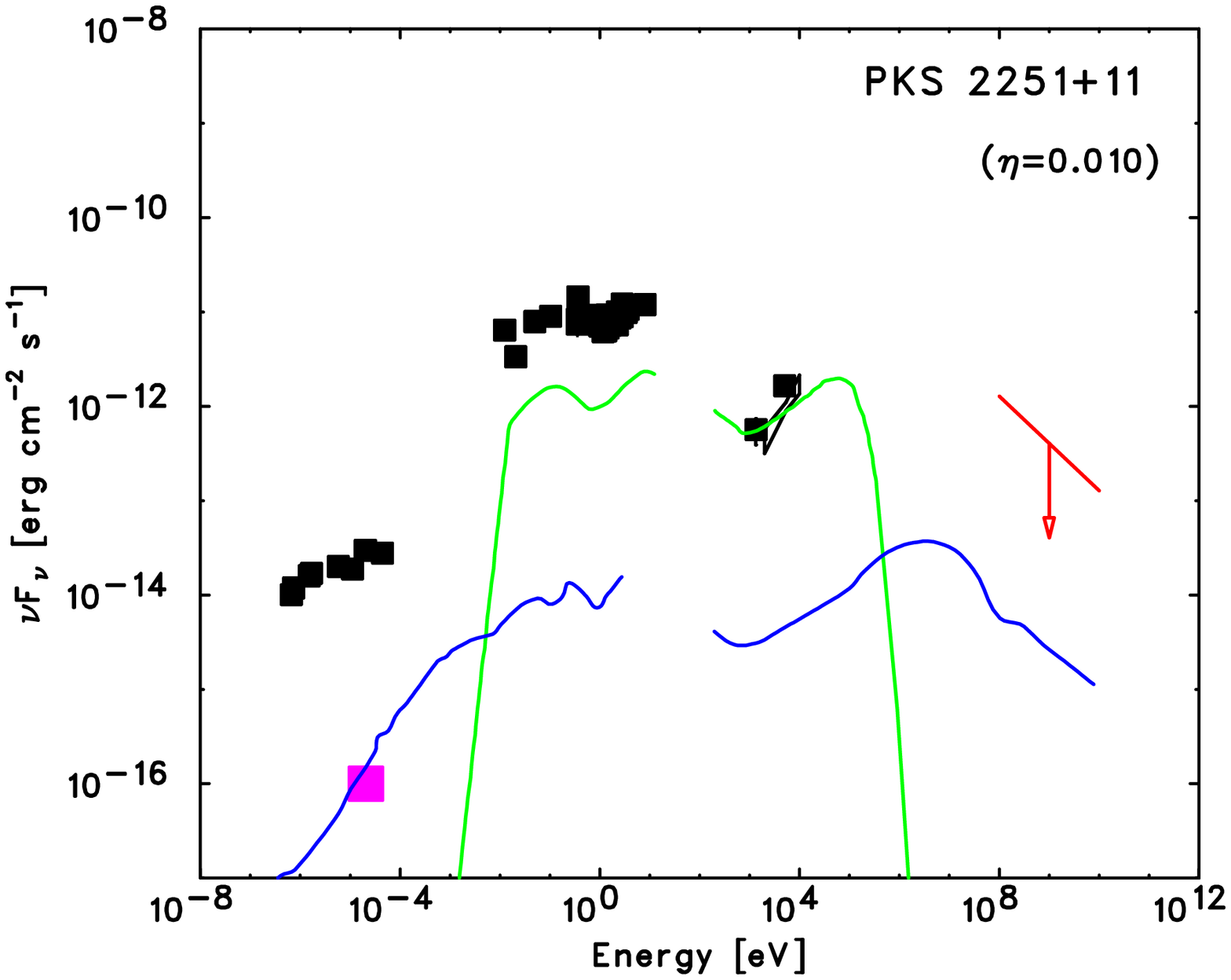}
\caption{$- continued$.}
\end{center}
\end{figure}

\begin{figure}
\begin{center}
\includegraphics[angle=0,scale=0.4]{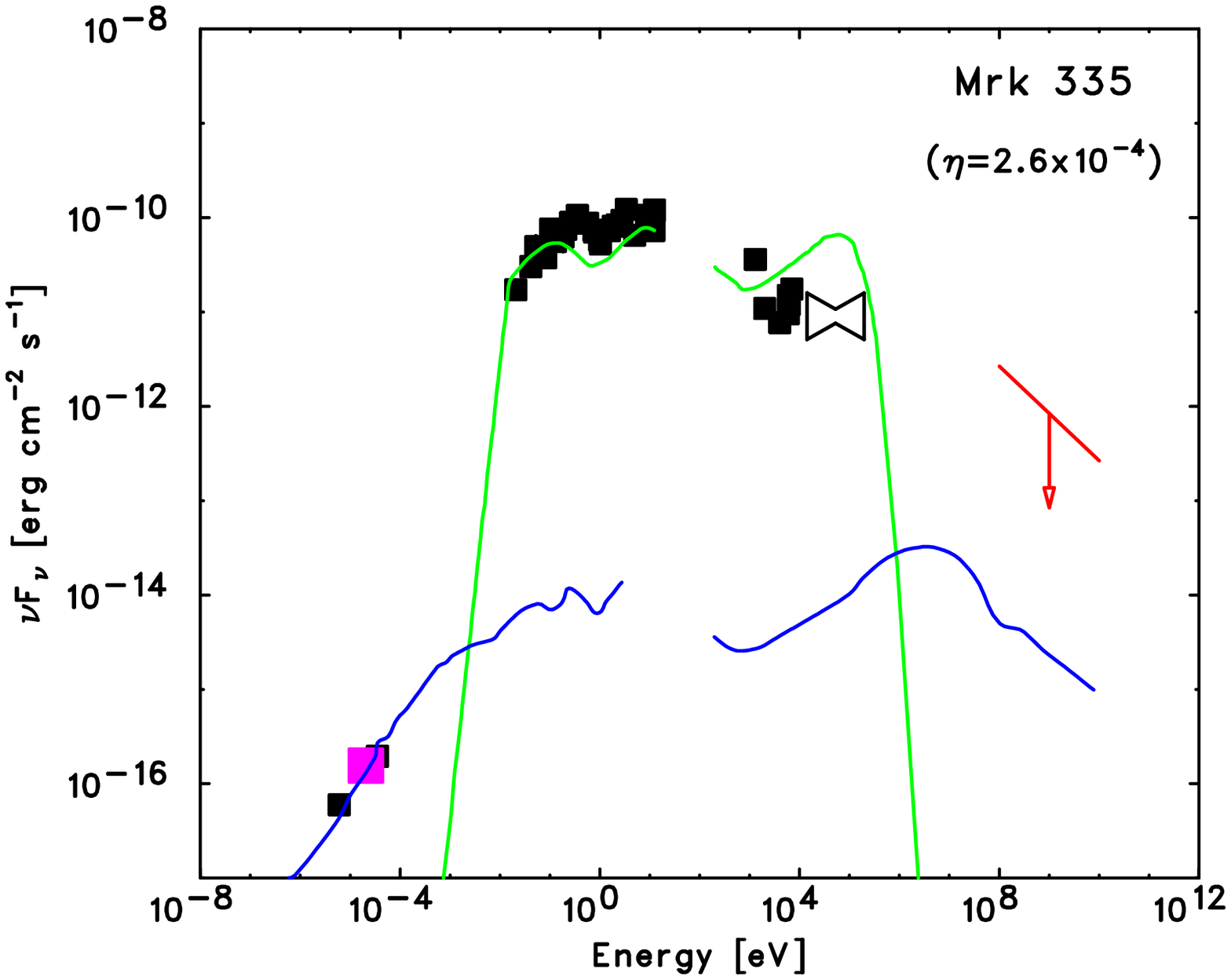}
\includegraphics[angle=0,scale=0.4]{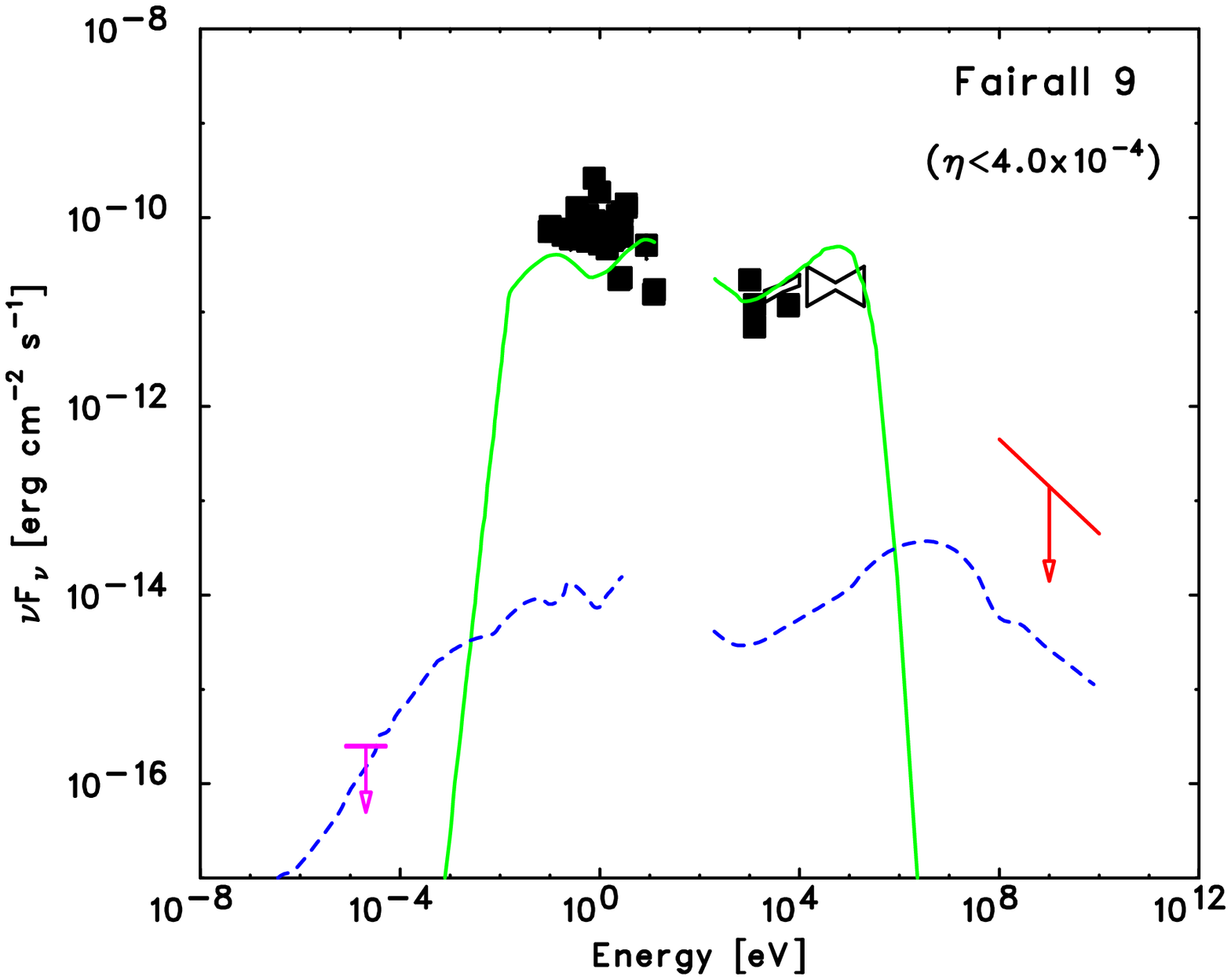}
\includegraphics[angle=0,scale=0.4]{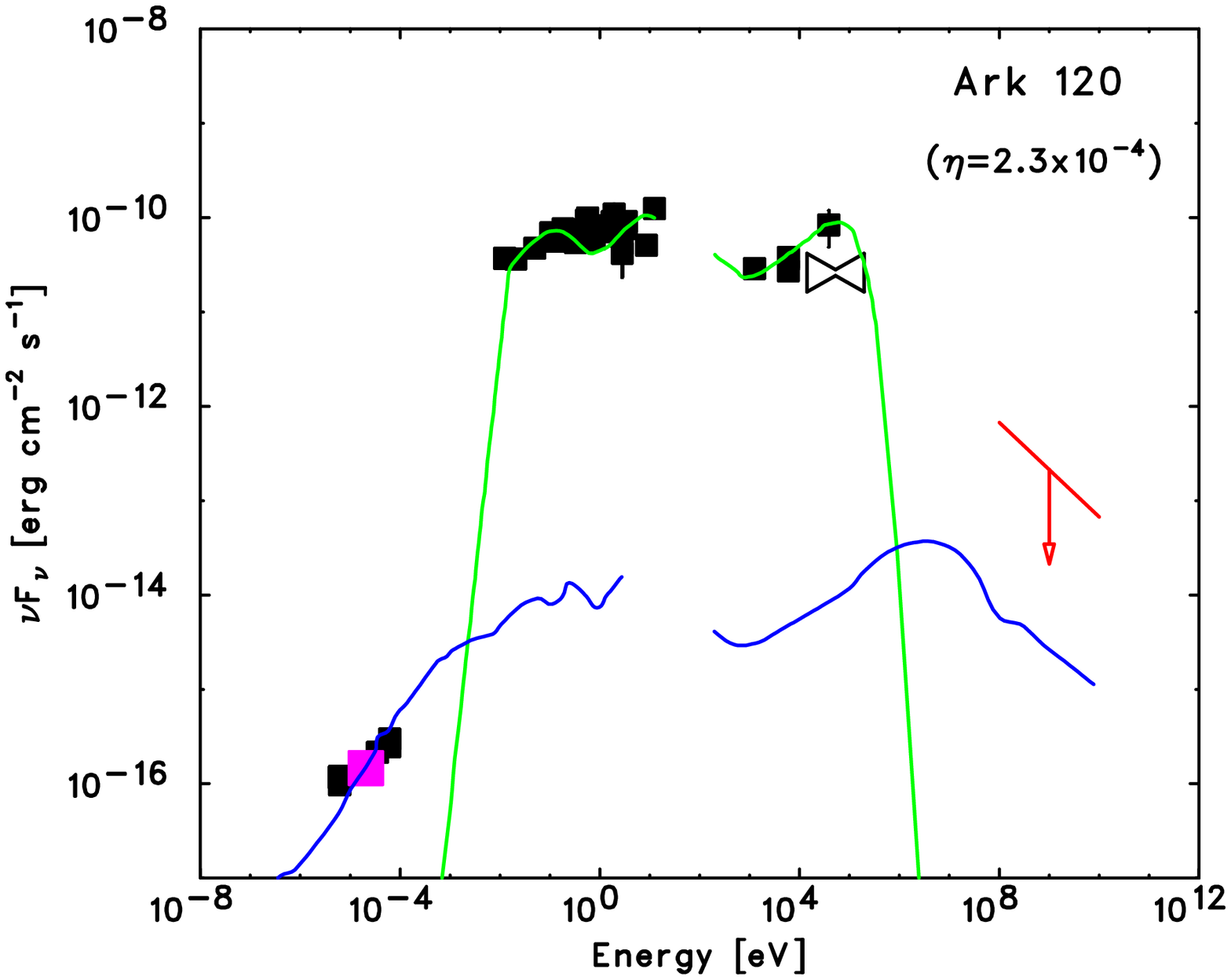}
\includegraphics[angle=0,scale=0.4]{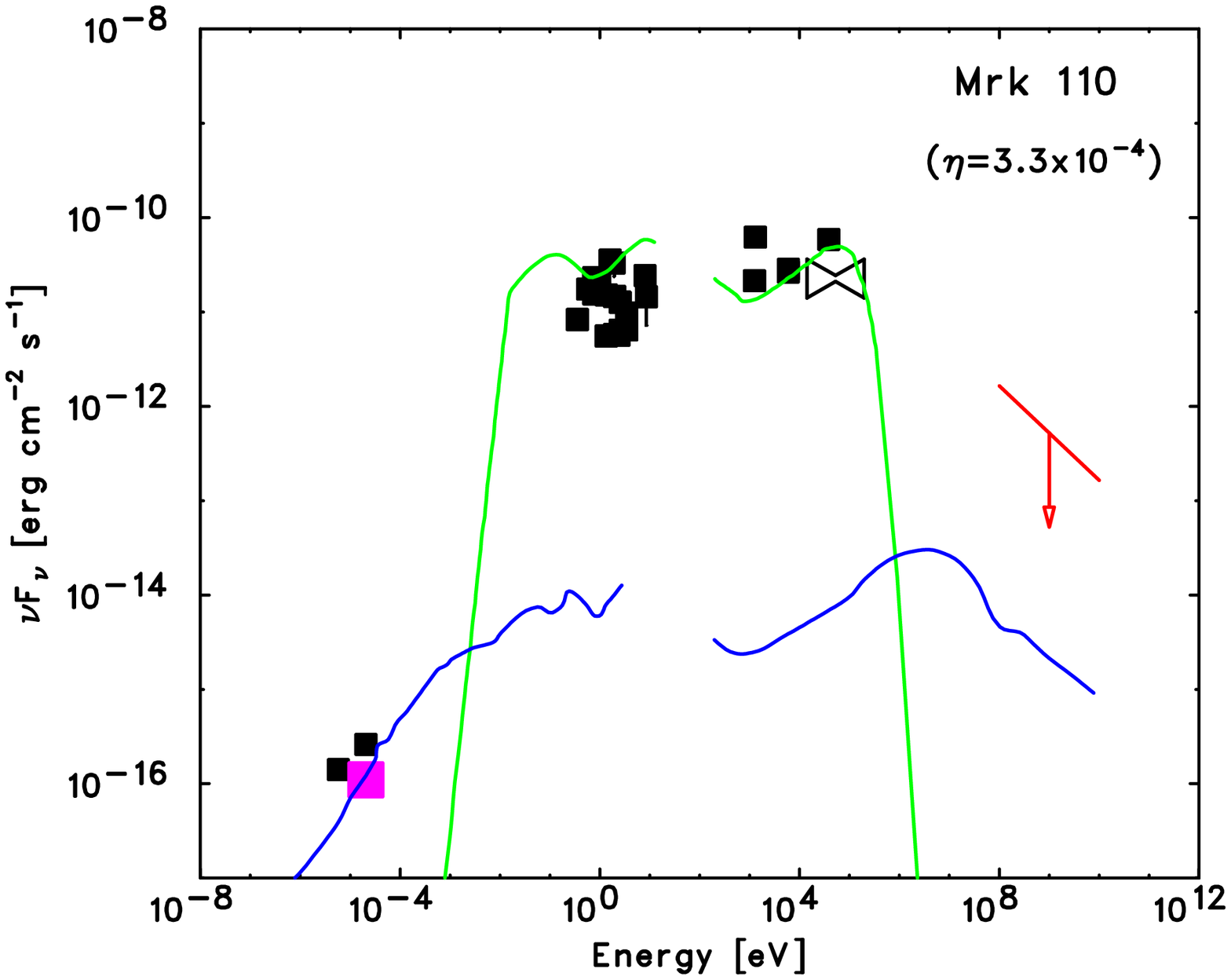}
\includegraphics[angle=0,scale=0.4]{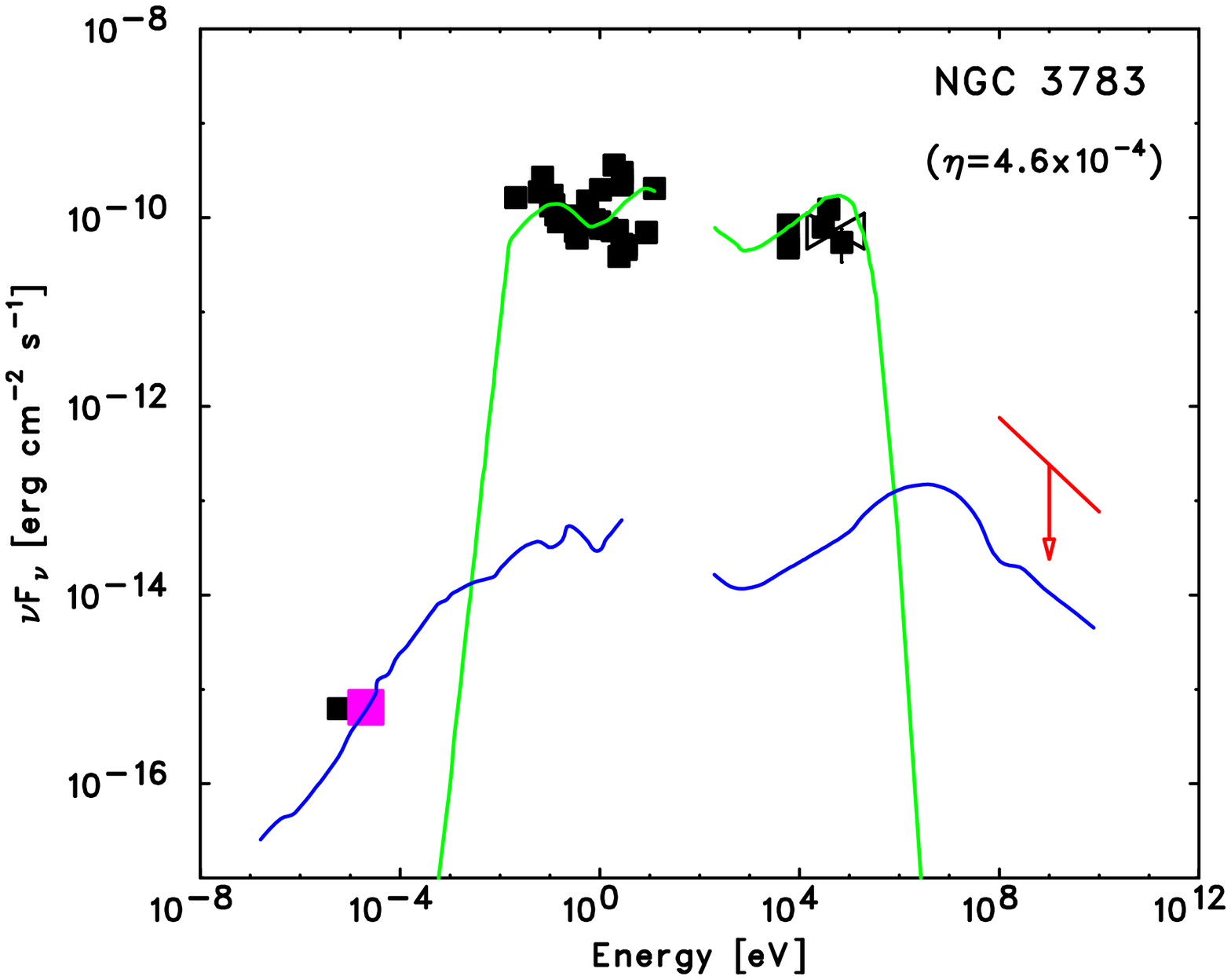}
\includegraphics[angle=0,scale=0.4]{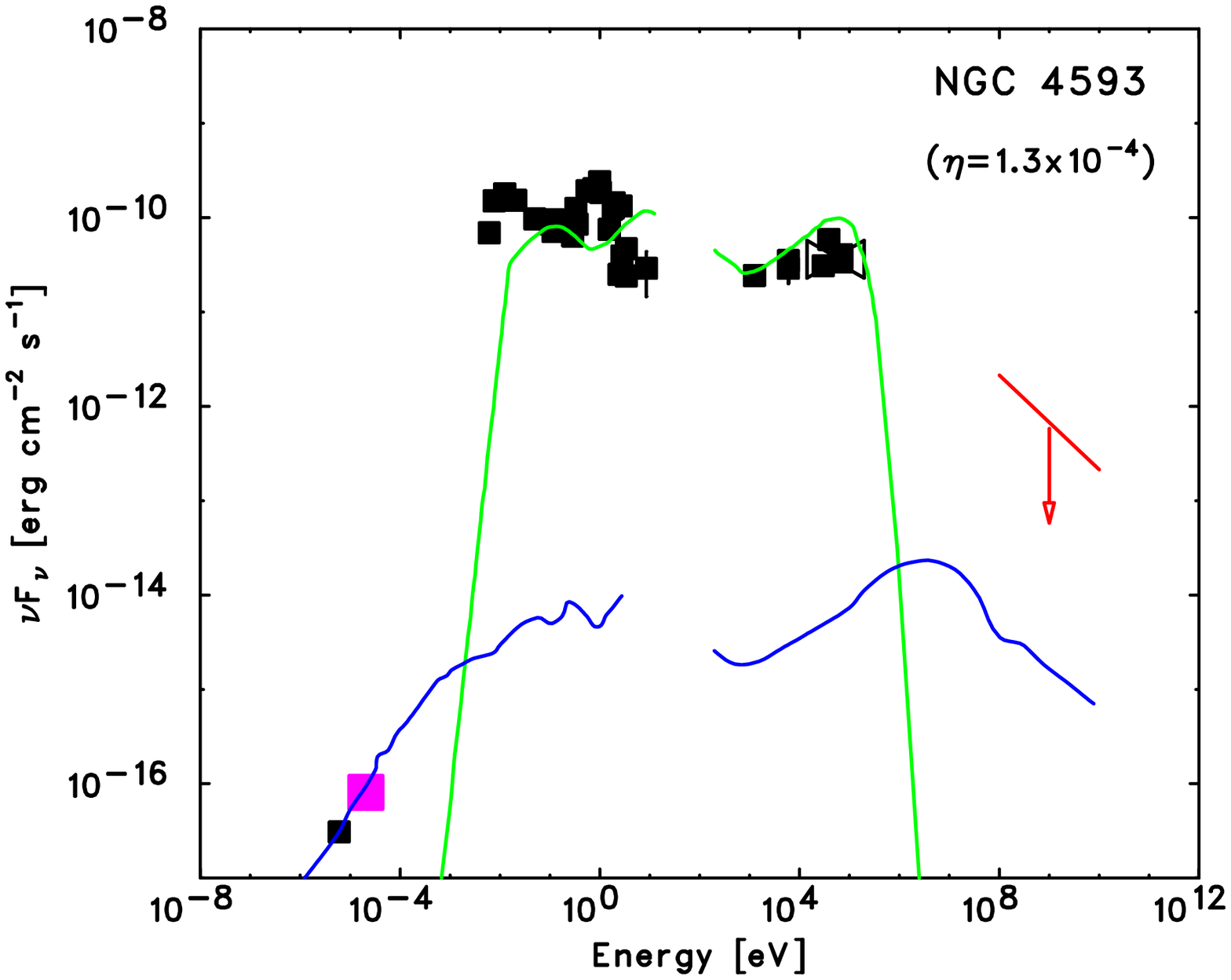}
\caption{Broad-band SEDs of the high-accretion Seyfert 1 galaxies. \F
 upper limits are indicated by red arrows. Black squares represent the
 historical data from NED. Magenta squares denote the $5$\,GHz radio
 fluxes of the unresolved nuclei, except for Fairall~9 where a radio
 detection has not been reported in the literature. 
The green curves correspond to the template of the accretion-related
 Seyfert-type emission \citep[from][]{kor99}, matched to the
 infrared--to--X-ray continuum of each source. The blue curves correspond
 to the broad-band spectrum of the quasar 3C~273 \citep[from][]{sol08},
 used here as a template of the jet-related emission and scaled to match
 the radio fluxes for each source. The mixing parameter $\eta$ for the
 phenomenological hybrid model discussed in \S\,4 is given in each
 panel.}
\end{center}
\end{figure}

\addtocounter{figure}{-1}
\begin{figure}
\begin{center}
\includegraphics[angle=0,scale=0.4]{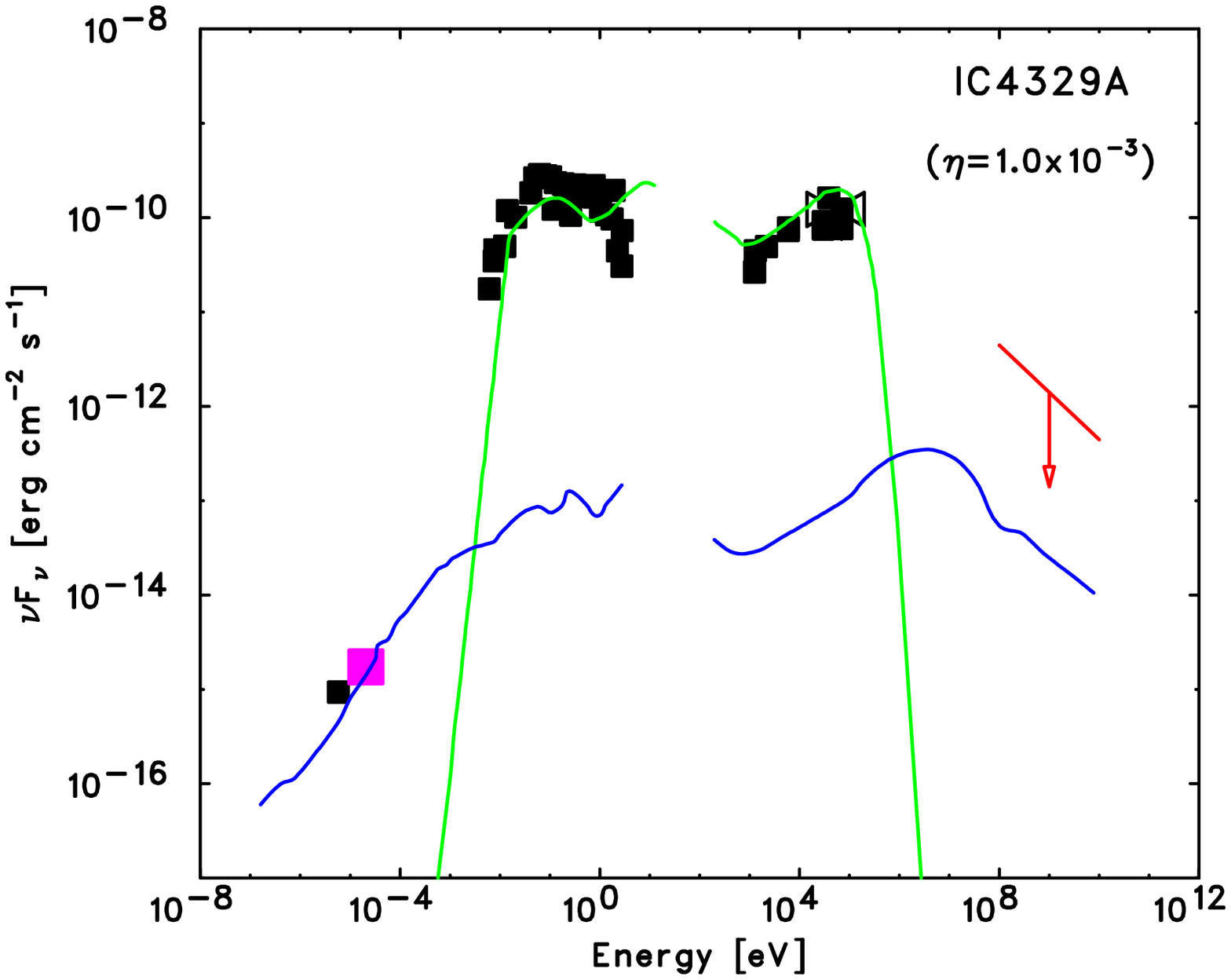}
\includegraphics[angle=0,scale=0.4]{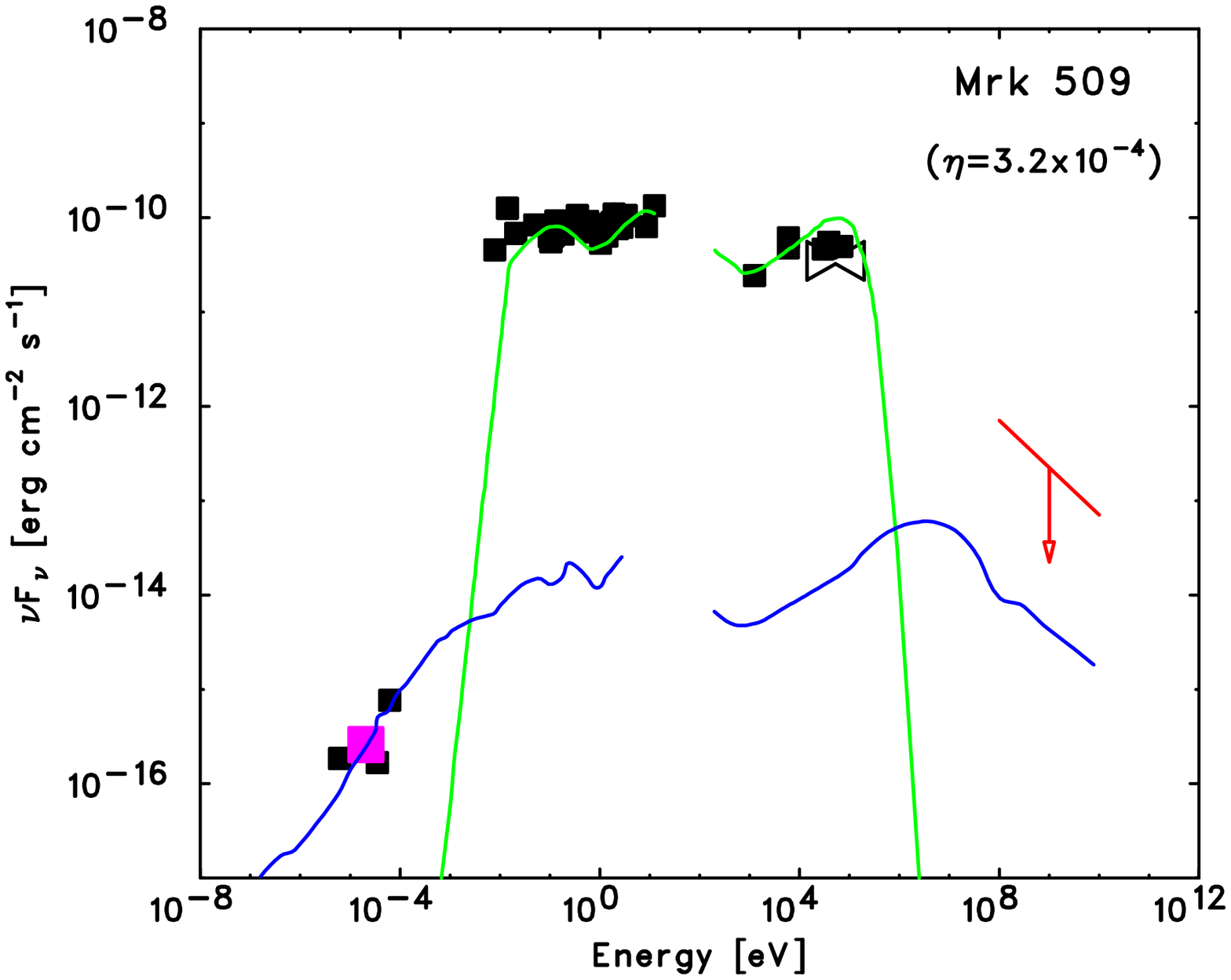}
\includegraphics[angle=0,scale=0.4]{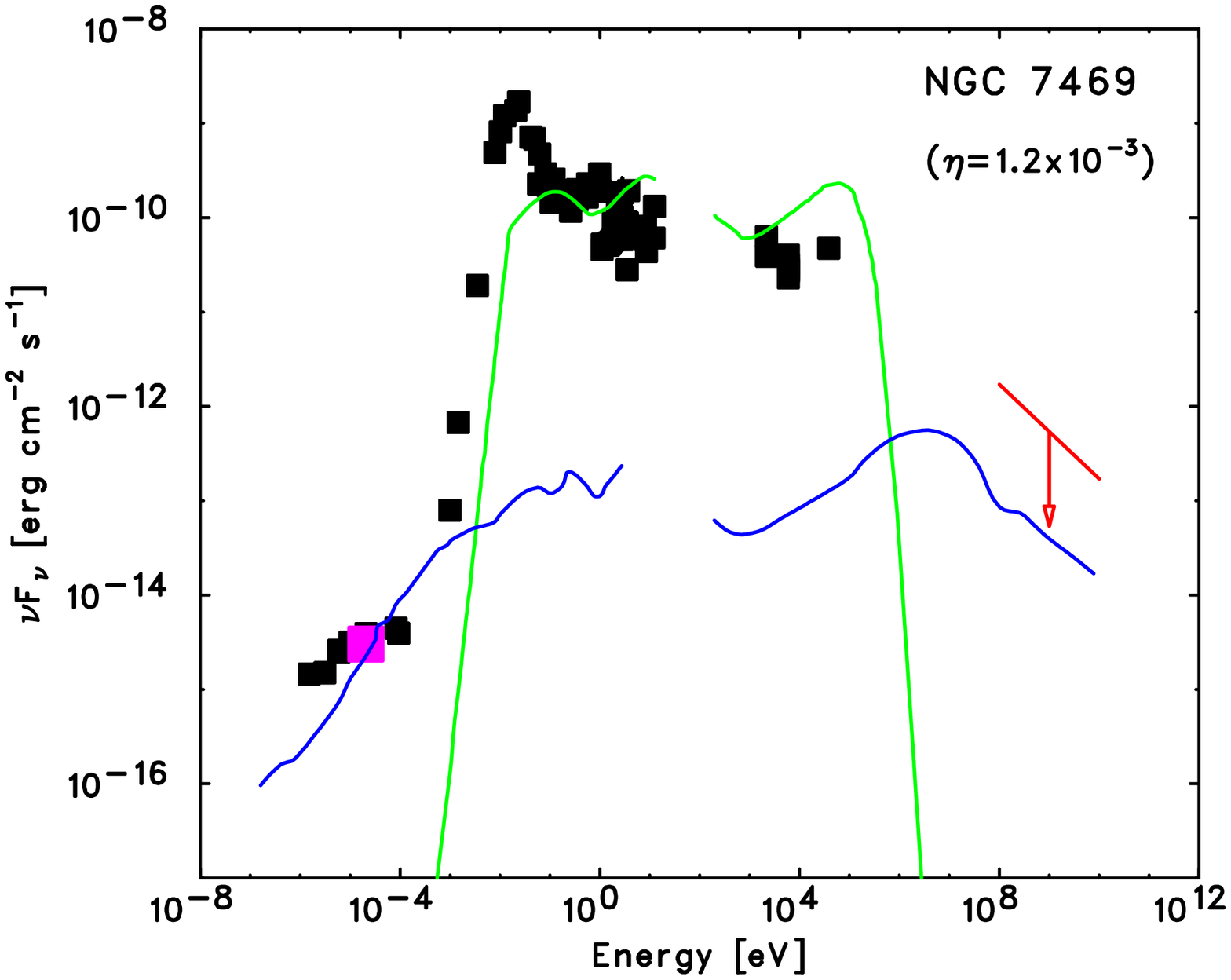}
\caption{$- continued$.}
\end{center}
\end{figure}

\end{document}